\documentclass{tlp}
\usepackage{amsmath}
\usepackage{alltt,amssymb,afterpage,fancyheadings,enumerate,wrapfig,rotating,theorem,float}
\input epsf
\newcommand{\boxps}[1]{\makebox{\epsfbox{#1.ps}}}

\newenvironment{alltts}{\begin{alltt}\def\baselinestretch{1.2}\footnotesize}{\end{alltt}}

\newfloat{program}{thbp}{lop}
\floatname{program}{Program}
\theoremstyle{break} 
\theorembodyfont{\sl} \newtheorem{Def}{Definition}[section]
\theorembodyfont{\sl} \newtheorem{The}{Theorem}[section]
\theorembodyfont{\sl} \newtheorem{Lem}{Lemma}[section]
\theoremheaderfont{\bf}

\title{Secure Prolog Based Mobile Code}

\author[S.W. Loke and A. Davison]
{SENG WAI LOKE\thanks{This author is grateful to Leon Sterling
who (with Andrew Davison) co-supervised the work on which this paper is based.}\\
DSTC, Monash University, \\
Caulfield, Victoria 3145, Australia\\
\email{swloke@dstc.monash.edu.au}
\and
ANDREW DAVISON\\
 Dept. of Computer Engineering \\   
Prince of Songkla University, \\
Hat Yai, Songkhla 90112, Thailand\\
\email{dandrew@ratree.psu.ac.th}}

\begin{document}

\maketitle
\label{firstpage}
\begin{abstract}
LogicWeb mobile code consists of Prolog-like rules embedded in Web pages,
thereby adding logic programming behaviour to those pages.  Since LogicWeb
programs are downloaded from foreign hosts and executed locally, there is
a need to protect the client from buggy or malicious code.  A security
model is crucial for making LogicWeb mobile code safe to execute. This
paper presents such a model, which supports programs of varying trust
levels by using different resource access policies. The
implementation of the model derives from an extended operational
semantics for the LogicWeb language, which provides a precise meaning
of {\em safety}.  
\end{abstract}

\section{Introduction}
A significant development of the World Wide Web, which has attracted
considerable interest in recent years, is the addition of {\em
mobile code} to Web pages; Java applets and
JavaScript are popular examples. Code of this kind
adds sophisticated interactive behaviour to Web documents, allowing more
useful tasks to be performed.  However, a key issue is security, since mobile
code is downloaded from foreign hosts and executed locally.

Malicious or buggy code can attack a local host in three 
main ways~\cite{brown96,olw97}:
\begin{itemize}
\item {\em integrity attacks}: attempts to delete or modify information in
the local environment in unauthorised ways. For example, 
a program issues an operating system command to delete local files. 

\item {\em privacy attacks}: attempts to steal or leak information to 
unauthorised parties. For example, a program reads a file 
and transmits the contents to another host.

\item {\em denial of service attacks}: attempts to occupy resources
to an extent which interferes with the normal operation of 
the host. An example is a program executing an infinite loop,
resulting in CPU cycles wasted and a depleted heap space.
\end{itemize}

These attacks are only possible if a foreign program has 
unregulated access to local system resources such as the file system,
network services (e.g., socket connections), CPU cycles, internal memory,
input/output devices, the program environment (e.g., environment variables),
and operating system commands. At the other extreme, a foreign program 
which is denied access to all resources cannot do any damage, but is
unlikely to be able to do much useful work. 

There is no one applicable policy for all foreign code: some
code should be more trusted than others, and hence should be given more
privileges. Providing a security model that provides the right level of
functionality with the appropriate degree of security at the right time
is a challenge. Indeed, security for mobile code is an active research
area~\cite{mos97,brown96,cgpv97,thorn97}.

An emerging trend is the use of formal language semantics to reason
with mobile code and prove security properties. For example,
Leroy and Rouaix~\shortcite{lr98} have developed formal techniques for
validating a typed functional language with assignments, ensuring
that memory locations always contain appropriate values. The use of a
rewriting logic is advocated in~\cite{mt97} as the basis for a mobile
code programming language where security properties can be formally
verified. Volpano ~\shortcite{volpano97} argues for the development of
provably-secure programming languages which can be remotely evaluated,
and whose semantics permit the formal proof of security properties.
As reported in~\cite{dfw96}, a number of security loopholes have been
discovered with Java applets, and this seems related to its lack of a
formal semantics. Recently, this weakness has been addressed by giving
semantics to subsets of Java~\cite{de98,jmt98}.

Declarative languages come with simple and elegant semantics,
providing a convenient basis for reasoning about program execution
and for introducing security mechanisms. This paper examines a
security extension for a Prolog-based mobile code language called the
LogicWeb language~\cite{ld98}. Underpinning the language is a view of the
Web as a collection of logic programs called {\em LogicWeb programs}.
LogicWeb clauses embedded within Web pages add logic programming behaviour
to those pages. A LogicWeb program can retrieve and manipulate other Web
pages (which may optionally contain embedded LogicWeb code themselves)
using compositional logic programming techniques. Applications of LogicWeb
include Web-based distributed deductive databases, and extending the
semantics of Web links~\cite{lds96a,lds96b,ld97a}.

In this paper, we present a security model for LogicWeb which
provides for the safe (i.e., host-protecting) 
execution of programs without overly restricting their capabilities.
The key features of the model are:
\begin{itemize}
\item {\em flexible}: it supports varying degrees of trust and access to
resources. More trusted programs can be given more privileges to achieve
grea\-ter functionality.  Policies determine what resources are available
to whom, and how the resources must be used.

\item {\em formally specified}: the model provides a precise meaning 
of {\em safety}. 

\item {\em implemented based on the model's specification}: 
the implementation takes the form of a meta-interpreter which is
closely allied to the underlying operational semantics.
\end{itemize}

The security model is specified as an extension of the operational
semantics of the LogicWeb language, and so permits a straightforward proof
that the model protects the host according to specified policies. We
describe how meta-programming provides the necessary control over
execution behaviour for a flexible yet safe execution environment,
including control over resource usage and looping. We also show how
security policies can support different trust levels, and how policies
can be combined to control the execution of an application composed from
programs with different trust levels. The security model is implemented
as an extension of the system described in~\cite{ld98b}.

The rest of this paper is organised as follows. \S2 provides background
on LogicWeb, \S3 gives an overview of the security model, \S4 describes
how we utilise digital signatures, and \S5 examines how security
policies are specified. The need to combine security policies during
goal evaluation is explained in \S6. \S7 describes how policies are
enforced by policy programs. An implementation of the security model
is given in \S8. Control of resource usage using meta-interpreters is
illustrated in \S9. \S10 reviews related work, and \S11 concludes.

\section{Preliminaries}
This section gives an overview of LogicWeb programs, the language,
and the architecture of the client-side LogicWeb system.

\subsection{LogicWeb Programs} 

LogicWeb programs are constructed from the data returned by HEAD,
GET or POST HTTP requests. When successful, a HEAD request returns
meta-information about a page, whereas GET and POST requests return
meta-information {\em and} Web pages.
 
\begin{sloppypar}
The LogicWeb program derived from a HEAD response consists of
two types of facts for holding the page's meta-information and URL:

\begin{itemize}
\item \texttt{about({\em FieldName}, {\em Value})}. \texttt{about/2}
facts store the page's meta-information as supplied by the server
(e.g., the content length). 

\item \texttt{actual\_url({\em URL})}. This holds the URL of the page
whose meta-information was retrieved from the server.  The actual URL
would be different from the URL used in the original HTTP request if
the request was redirected by the server.

\end{itemize}
\end{sloppypar}

\begin{sloppypar}
The program corresponding to a HEAD response uses the identifier
\texttt{lw(head, URL)}. Programs corresponding to GET and POST responses
use \texttt{lw(get, URL)} and \texttt{lw(post(Data), URL)} 
respectively. A GET or POST response is translated into five types
of facts based on the meta-information returned by the page's server
and the HTML text in the page. Additional LogicWeb clauses may also be
included within the page. 
\end{sloppypar}

The facts include \texttt{about/2} and \texttt{actual\_url/1}
described above, and the following:
\begin{itemize}
\item \texttt{my\_id({\em Type}, {\em URL})}. \texttt{\em Type}
stores the type of the program, which is either the term \texttt{get}
or \texttt{post({\em Data})}. \texttt{\em Data} is the information
posted to the CGI script at \texttt{\em URL}.

\item \texttt{h\_text({\em HTMLText})}. This contains the
HTML text of the page (apart from any additional LogicWeb clauses).

\item \texttt{link({\em Label}, {\em URL})}. This stores
a page's link information. For instance, the anchor:
\begin{alltts}
...<A HREF="http://www.cs.mu.oz.au/">Melb. U</A>...
\end{alltts}
becomes:
\begin{alltts}
link("Melb. U", "http://www.cs.mu.oz.au/").
\end{alltts}

\end{itemize}

Additional facts can be readily extracted from the text of
the page, as described in~\cite{ld98b}.

\begin{sloppypar}
Clauses are included within a page between the tags ``{\tt <LW\_CODE>}''
and ``{\tt </LW\_CODE>}''. Typically, the code appears inside a
verbatim container ``{\tt <PRE>...</PRE>}'' or a comment container
``{\tt <!--...-->}'' so that it is uninterpreted by the browser.
Figure~1 shows a page holding LogicWeb clauses about research interests.
\end{sloppypar}

\begin{figure}[h]
\begin{center} 
\begin{alltts}
\def\baselinestretch{1.0} 
{\footnotesize
<HTML>
<HEAD>
<TITLE>Seng Wai Loke's Home Page</TITLE>
</HEAD>

<BODY><H1>Seng Wai Loke's Home Page</H1>
I'm from the<A HREF="http://www.cs.mu.oz.au/">
Department of Computer Science</A> at the 
<A HREF="http://www.unimelb.edu.au/">
University of Melbourne</A>.
<!--
<LW_CODE>
interests(["Logic Programming", "AI", "Web", "Agents"]).

friend_page("http://www.cs.mu.oz.au/~f1/").
friend_page("http://www.cs.mu.oz.au/~f2").

interested_in(X) :- interests(Is), member(X, Is).
interested_in(X) :-
  friend_page(URL), lw(get, URL)#>interested_in(X).
</LW_CODE>
-->
</BODY>
</HTML>
}

\end{alltts}
\caption{A page with clauses describing research interests.}
\label{fig:3_3lwc}
\end{center}
\end{figure}

\begin{sloppypar}
\texttt{interested\_in/1} defines the author's interests using
Prolog-style facts and rules. The LogicWeb operator
``\texttt{\#>}'' evaluates the \texttt{interested\_in/1} goal against
the program identified by \texttt{lw(get, URL)}.
\end{sloppypar}

\begin{sloppypar}
Assuming that the page shown in Figure~1 was retrieved 
from \texttt{http://www.cs.mu.oz.au/\~{}swl/} using the
GET method, then the resulting LogicWeb program contains:

\begin{enumerate}
\item Several \texttt{about/2} facts concerning the page's 
meta-information.

\item \texttt{actual\_url("http://www.cs.mu.oz.au/\~{}swl/")}.

\item \texttt{my\_id(get,"http://www.cs.mu.oz.au/\~{}swl/")}.

\item \texttt{h\_text("<HTML>...</HTML>")}. This fact stores
all the HTML text from the page except the clauses between 
the ``{\tt <LW\_CODE>}'' and ``{\tt </LW\_CODE>}'' tags.

\item Two \texttt{link/2} facts:
\begin{alltts}
link("Department of Computer Science", "http://www.cs.mu.oz.au/").
link("University of Melbourne", "http://www.unimelb.edu.au/").
\end{alltts}

\item The clauses between the ``{\tt <LW\_CODE>}'' and
``{\tt </LW\_CODE>}'' tags.
\end{enumerate}
\end{sloppypar}

A {\em LogicWeb application} comprises a set of LogicWeb programs,
with one singled out (by the programmer) as the {\em main program}.
Each application has an {\em identity} which is the program identifier
of the main program, and a {\em visible interface} with which the user
interacts. A typical interface consists of a HTML form and/or Web
links. Clauses in the main program define the forms interface and the
mapping of link selections to goals.

\subsection{The LogicWeb Language}

The LogicWeb language utilises Prolog and special LogicWeb 
operators. We will assume that the reader is familiar with Prolog,
and concentrate instead on the operators.

\subsubsection{LogicWeb Operators}

A {\em LogicWeb goal} is formed using the operator ``\texttt{\#>}'',
which is known as {\em context switching}. For example, the LogicWeb 
goal \texttt{lw(get, URL)\#>Goal} applies \texttt{Goal} to the program 
specified by the \texttt{lw(get, URL)} identifier.

If the program is not already present on the client-side, then its page
will be downloaded and translated into a LogicWeb program before the
query is evaluated. However, if the program is present, then the goal is
executed immediately. Thus, the ``\texttt{\#>}'' operator permits the
programmer to think of Web computation as goals applied to programs,
with no need for explicit Web page retrieval, parsing, or storage.

\begin{sloppypar}
The current context of a LogicWeb goal, i.e. the program (or composition 
of programs) where the goal is located, is ignored when ``\texttt{\#>}''
is evaluated.
\end{sloppypar}

LogicWeb programs can be manipulated using five {\em LW-composition
operators}. Four are based on Brogi's algebraic program
composition framework~\cite{brogi93}, and the fifth on 
{\em reduce} from functional programming. The operators are:
{\em LW-union} (``\texttt{+}''), 
{\em LW-intersection} (``\texttt{*}''), 
{\em LW-restriction} (``\texttt{/}''),
{\em LW-en\-cap\-su\-la\-tion} (``\texttt{@}''), and {\em LW-reduce}
(``\texttt{<>}''). An expression formed with LW-composition
operators is called a {\em program expression}.

LogicWeb programs are composed together after they have been retrieved
from the Web, which means that the semantics of the operators can be
viewed as a variant of Brogi's definitions, extended to address issues
related to page downloading and the translation of pages to programs.
The operational semantics are specified in \S2.2.3.

\begin{sloppypar}
The sixth operator, the {\em context} operator, denoted by 
``\texttt{(\#)}'', represents the current context in a program expression.
For instance:
\begin{alltts}
?- lw(get, "URL0")#>(((#) + lw(get, "URL1"))#>interested_in(X)).
\end{alltts}
``\texttt{(\#)}'' is instantiated to \texttt{lw(get, "URL0")}
when the goal is evaluated. ``\texttt{(\#)}'' can be used in place
of a program identifier in any expression, which provides very useful 
expressive power. For instance, it can readily model
other composition formalisms~\cite{ld98}.
\end{sloppypar}

\subsubsection{EBNF Syntax} 
A {\em pure} LogicWeb program is a finite set of clauses of the form
 $[\forall x](\mathcal{A}~\texttt{:-}~\mathcal{G})$
where $\mathcal{G}$ is defined recursively as:
\begin{equation*}
 \mathcal{G}  ::=  \mathcal{A}~|~\mathcal{E}~\text{\tt \#>}~\mathcal{G}
~|~(\mathcal{G},\mathcal{G}) 
\end{equation*}
 
$\mathcal{E}$ defines LogicWeb program expressions:
\begin{align*}
 \mathcal{E}  ::= ~&  \mathcal{P}~|~\mathcal{E} \texttt{+} \mathcal{E}
~|~\mathcal{E} \texttt{*} \mathcal{E}
~|~\mathcal{E} \texttt{/} \mathcal{P}~|~\texttt{@}E
~|~\texttt{(/)<>($\mathcal{E}$,$\mathcal{L}_{(\mathcal{P})}$)}
~|~\texttt{($\oplus$)<>$\mathcal{L}_{(\mathcal{E})}$}\\
 \mathcal{P}  ::= ~& \text{\tt lw(head,}~URL\text{\tt )}
~|~\text{\tt lw(get,}~URL\text{\tt )}
~|~\text{\tt lw(post(}\mathcal{L}_{(\mathcal{F})}\texttt{),}~URL\texttt{)}
~|~\texttt{(\#)}\\
 \mathcal{F} ::= ~&\texttt{field($Name$,$Value$)} \\
 \mathcal{L}_{(\mathcal{I})} ::= ~&\text{\tt []}
~|~\texttt{[$\mathcal{I}$|$\mathcal{L}_{(\mathcal{I})}$]}\\
\mathcal{\oplus} ::= ~& \texttt{+}~|~\texttt{*} 
\end{align*}
$\mathcal{L}_{(\mathcal{I})}$ defines a Prolog list of items, each of
which is described by a nonterminal $\mathcal{I}$. $URL$ is a URL, and
$\mathcal{F}$ a query attribute submitted to a CGI script. $Name$
is the name of a query attribute, and $Value$ is the value submitted
to the server for the corresponding attribute. The context operator
``\texttt{(\#)}'' can appear anywhere a program identifier can, but
must be instantiated to a program identifier when used
as the right-hand argument of LW-restriction.

\subsubsection{Operational Semantics}
The operational semantics of the LogicWeb language is detailed
in~\cite{ld98}, but so that this paper is self-contained, the semantics
are summarised here. In particular, we do not consider 
LW-restriction and LW-reduce.

\medskip

An {\em oracle function} is defined to
model the accessing of LogicWeb programs. 

\begin{Def}[oracle function] 
The oracle function 
\[ download: LWProgramIDs \rightarrow LWPrograms \cup \{\bot\}\]
takes a LogicWeb program identifier $P$ (of the form
defined by $\mathcal{P}$), and returns
the program $D_P$ if it is successfully created. 
Failure to obtain a program is represented 
by returning the symbol $\bot$.
\begin{equation*}
  download(P) =
 \begin{cases} 
 D_P & \text{if the program denoted by identifier $P$ is} \\
     & \text{successfully created,} \\
 \bot & \text{otherwise.}
\end{cases}
\end{equation*}
\end{Def}

$download$ attempts to download a HTTP response object and translate
it into a LogicWeb program in the way specified in \S2.1. 

\medskip

\begin{sloppypar}
The set of downloaded programs is extended by calling the function
$add\_programs$, which takes the set of existing LogicWeb
programs and maps it to a new set using $download$.
\end{sloppypar}
\begin{sloppypar}

\begin{Def}[addition of LogicWeb programs] 
Let $\wp$ denote powerset.
The function \[add\_programs: \wp (LWPrograms) \times 
\wp (LWProgramIDs) \rightarrow \wp (LWPrograms)\]
takes a set $S$ of programs and a set $I$ of program identifiers
and returns a new set
$add\_programs(S,I)$ consisting of $S$ augmented with newly created 
programs mentioned in $I$ but previously not in $S$:
\begin{equation*}
add\_programs(S,I) = S \cup \{ D_P~|~P \in (I~\backslash~ids(S)),
~D_P = download(P),~D_P \not= \bot \}
\end{equation*} 
where $ids$ is a function that takes a set of programs and returns
the identifiers of the programs in the set, i.e.
$ids(S)$ is the set of identifiers of the programs in $S$.

\end{Def}
\end{sloppypar}

\medskip

A derivation relation involving the set of downloaded LogicWeb programs
specifies the computation model involving interactions with the Web.

\begin{Def}[derivation relation]
For any goal formula $G$ and program expression $E$, we denote by $S,
E \vdash^{S'}_{\theta} G$ the fact that there exists a {\em top-down
derivation} of $G$ in $E$ starting with the set $S$ of existing
LogicWeb programs and ending with computed answer substitution $\theta$
and created program set $S'$. A top-down derivation or proof of $G$
in $E$ starting with $S$ and ending with $S'$ and $\theta$ is a finite tree
such that:
\begin{enumerate}
\item the root node (bottom node) is labelled by the string
``$S, E \vdash^{S'}_{\theta} G$'';

\item the internal nodes are derived according to the set of 
inference rules given below; and

\item all the leaves of the tree are either empty or labelled by a 
string not containing the symbol ``$\vdash$'' (e.g., 
the label ``$(A~\texttt{:-}~G) \in P$'').
\end{enumerate} 
$\vdash_{\theta}^{S}$ is defined to be the smallest relation satisfying
the inference rules below. If $S, E \vdash^{S'}_{\theta} G$, then the
goal $G$ $succeeded$ when evaluated in $E$ using $S$. Otherwise,
the goal $G$ $failed$ when evaluated in $E$ using $S$. 

\end{Def}

Given a top-down derivation $S, E \vdash^{S'}_{\theta} G$, if the
derivation does not involve interactions with the Web or if no programs
are downloaded, then $S'$ is the same as $S$ (i.e., $S' = S$). If
programs are downloaded, then $S$ is extended to $S'$ ($\supseteq S$). The
difference $S'~\backslash~S$ represents the effect of Web interactions
during the derivation.

\medskip

A context is defined for each node in a top-down derivation whose label
is of the form ``$S, E \vdash^{S'}_{\theta} G$''.

\begin{Def}[context of a goal]
Given the node label ``$S, E \vdash^{S'}_{\theta} G$'',
$E$ is the $context$ (of $G$). 
\end{Def}

In the rules, $P$ denotes single program identifiers of the form
$\mathcal{P}$, and $E$ and $F$ denote program expressions of the form
$\mathcal{E}$ as defined in \S2.2.2. $L_{(\mathcal{I})}$ denotes a list
of the form $\mathcal{L}_{\mathcal{I}}$. $\epsilon$ denotes the empty
(identity) substitution.

The following rules
define derivation in pure Prolog taking into account 
the creation of LogicWeb programs: 
\paragraph{\bf True.}
\begin{equation}
    \frac{\;}{S, E \vdash_{\epsilon}^{S} \texttt{true}}
\end{equation}
\texttt{true} 
is always derivable in any program expression $E$ without any
change to the set of created programs.

\paragraph{\bf Conjunction.}
\begin{equation}
    \frac{S, E \vdash_{\theta}^{S'} G_1\;\;\;\wedge\;\;\;S', E \vdash_{\gamma}^{S''} G_2\theta}{S, E \vdash_{\theta \gamma}^{S''} G_1,G_2}
\end{equation}
$S'$, which may or may not be the same as $S$,
is the result of Web interactions from $G_1$'s derivation.
The result of these interactions are propagated to $G_2$ by
starting the derivation of $G_2\theta$ with $S'$. Since $G_2\theta$
started with $S'$, $S''$
is the result of Web interactions during the derivation of
the conjunction.

\paragraph{\bf Atomic formula.}
\begin{equation}
 \frac{S, E \vdash_{\theta}^{S'} (H ~\texttt{:-}~ G)\;\;\;\wedge\;\;\;
\gamma=mgu(A,H\theta)\;\;\;\wedge\;\;\;
S',E \vdash_{\delta}^{S''} G\theta\gamma}{S, E \vdash_{\theta \gamma \delta}^{S''} A}
\end{equation} 
Obtaining clauses from $E$ can involve the creation of new programs due
to LW-\-en\-cap\-su\-la\-tion (see rule (8)), and so, $S$ is changed
to $S'$. The proof of the body starts with the computed program set $S'$
and returns the new set $S''$ and the answer substitution $\delta$.

\paragraph{\bf Obtaining clauses from a single program.}
\begin{equation}
\frac{(A ~\texttt{:-}~G) \in P}
{S, P \vdash_{\epsilon}^S (A ~\texttt{:-}~ G)}
\end{equation}
The answer substitution is $\epsilon$, and there is no change to $S$. 

\medskip

\begin{sloppypar}
Below, we present the rules for choosing clauses from LW-union,
LW-intersection, LW-encapsulation, and context switching. 
\end{sloppypar}

\paragraph{\bf LW-union.}
\begin{equation}
\frac{S, E \vdash_{\theta}^{S'} (A ~\texttt{:-}~G)}
{S, E \texttt{+} F \vdash_{\theta}^{S'} (A ~\texttt{:-}~G)}
\end{equation}
\begin{equation}
\frac{S, F \vdash_{\theta}^{S'} (A ~\texttt{:-}~G)}
{S, E \texttt{+} F \vdash_{\theta}^{S'} (A ~\texttt{:-}~G)}
\end{equation}
A clause is chosen from a LW-union $E \texttt{+} F$ by choosing
a clause from either $E$ or $F$.

\paragraph{\bf LW-intersection.}
\begin{equation}
\frac{S, E \vdash_{\theta_1}^{S'} (H_1 ~\texttt{:-}~ G_1)
\;\;\;\wedge\;\;\;
S', F \vdash_{\theta_2}^{S''} (H_2 ~\texttt{:-}~ G_2)
\;\;\;\wedge\;\;\;
\gamma=mgu(H_1\theta_1,H_2\theta_2)}
{S, E \texttt{*} F \vdash_{\theta_1 \theta_2 \gamma}^{S''} 
(H_1 ~\texttt{:-}~G_1,G_2)}
\end{equation} 
A clause $H \texttt{:-} G$ is obtained from 
the LW-intersection $E \texttt{*} F$ if there exists a clause
$H_1 \texttt{:-} G_1$ in $E$ and a clause $H_2 \texttt{:-} G_2$ in $F$ 
such that $H$ unifies with $H_1$ and $H_2$,
and $G=(G_1,G_2)$. 
This rule utilises a left-to-right ordering in choosing clauses from
$E \texttt{*} F$. Rules are first chosen from $E$ returning $S'$,
and then, $S'$ is used when selecting clauses from $F$ 
ending up with $S''$. 

\paragraph{\bf LW-encapsulation.}
\begin{equation}
\frac{S, E \vdash_{\theta}^{S'} A}
{S, \texttt{@}E \vdash_{\theta}^{S'} A ~\texttt{:-}~\texttt{true}}
\end{equation}
A clause $A \texttt{:- true}$ is obtained from the encapsulation of
$E$, $\texttt{@}E$, if $A$ is provable in $E$. 
LogicWeb programs can be created in the proof of $A$, i.e.
a new program set $S'$ is computed starting with $S$.

\paragraph{\bf Context Switching.}
We first define the function
\[expids: ProgramExpressions \rightarrow \wp (LWProgramIDs)\] 
to refer to the program identifiers within a program expression. 
$expids$ is defined recursively based 
on the syntax of program expressions:
\begin{align*}
expids(P) =~& \{P\} \\
expids(E_1~\texttt{+}~E_2) =~ & expids(E_1) \cup expids(E_2)\\
expids(E_1~\texttt{*}~E_2) =~ & expids(E_1) \cup expids(E_2)\\
expids(E~\texttt{/}~P) =~ & expids(E) \cup expids(P)\\
expids(\texttt{@}E) =~ & expids(E) \\
expids(\texttt{(/)<>($E$,$L_{(\mathcal{P})}$)}) = ~& 
expids(E) \cup  expids(L_{(\mathcal{P})})\\
expids(\texttt{($\oplus$)<>$L_{(\mathcal{E})}$}) = ~&  
expids(L_{(\mathcal{E})})\\
expids(L_{(\mathcal{E})}) = ~& 
\bigcup_{E \in L_{(\mathcal{E})}} expids(E)
\end{align*}
In the above, we have used $\in$ to represent list membership, and a 
  $L_{(\mathcal{P})}$ is a $L_{(\mathcal{E})}$ from the EBNF 
definition in \S2.2.2.

We also define the function:
\[insertCC: ProgramExpressions \times ProgramExpressions 
\rightarrow ProgramExpressions\]
which substitutes every occurrence of the operator \texttt{(\#)}
in a program expression (the first argument)
with the current context (the second argument):
\begin{align*}
insertCC(\texttt{(\#)}, C) = ~& C \\
insertCC(P, C) =~& P \\
insertCC(E_1~\texttt{+}~E_2,C) =~ & 
insertCC(E_1,C)~\texttt{+}~insertCC(E_2,C)\\
insertCC(E_1~\texttt{*}~E_2,C) =~ & 
insertCC(E_1,C)~\texttt{*}~insertCC(E_2,C)\\
insertCC(E~\texttt{/}~P,C) =~ & insertCC(E,C)~\texttt{/}~insertCC(P,C)\\
insertCC(\texttt{@}E,C) =~ & \texttt{@}insertCC(E,C) \\
insertCC(\texttt{(/)<>($E$,$L_{(\mathcal{P})}$)},C) = ~& 
\texttt{(/)<>($insertCC(E,C)$,$insertCC(L_{(\mathcal{P})},C)$)} \\
insertCC(\texttt{($\oplus$)<>$L_{(\mathcal{E})}$},C) = ~& 
\texttt{($\oplus$)<>$insertCC(L_{(\mathcal{E})},C)$}\\
insertCC(L_{(\mathcal{E})},C) = ~& 
[insertCC(E, C)~|~E \in L_{(\mathcal{E})}]
\end{align*}

The rule defining context switching is the following:
\begin{equation}
\frac{I \subseteq ids(S')\;\;\;\wedge\;\;\; 
S', F' \vdash_{\theta}^{S''} G}
{S, E \vdash_{\theta}^{S''} F~\texttt{\#>}~G}
\end{equation}
where $F' = insertCC(F, E)$, $I = expids(F)$, and
$S' = add\_programs(S,I)$.

The rule states that the goal $F~\texttt{\#>}~G$ is provable in $E$
starting with the program set $S$ if the goal $G$ is provable in $F'$
starting with the updated program set $S'$ which contains all the programs
mentioned in $I$ (and hence, in $F$).

\subsection{The LogicWeb System Architecture} 

The architecture of the LogicWeb system (its components and 
data-flows) is shown in Figure~2.

\begin{figure}[H]
\begin{center} 
\setlength{\epsfxsize}{4.4in} 
\boxps{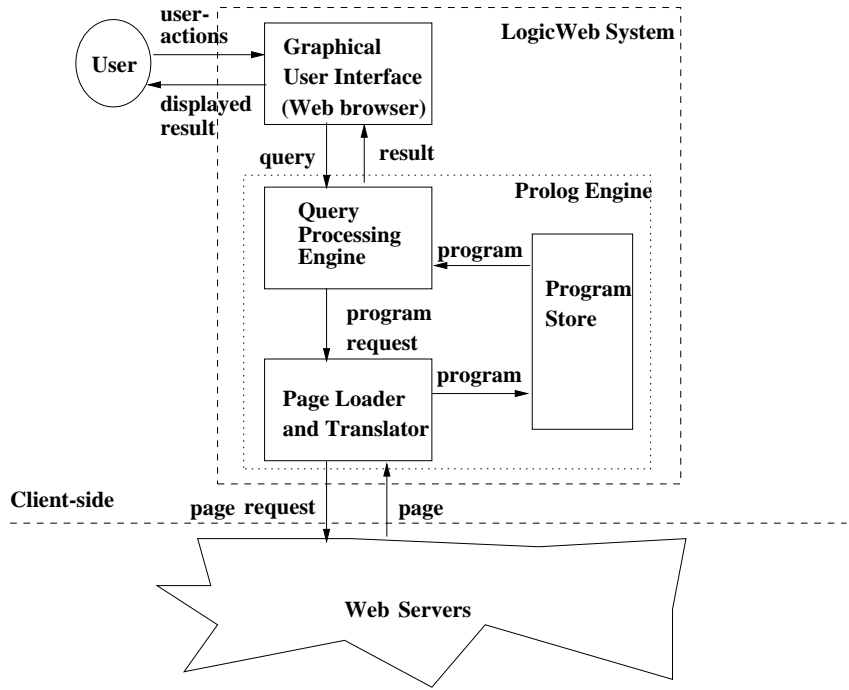}
\caption{Architecture of the LogicWeb system.}
\label{fig:3_6lwarch}
\end{center}
\end{figure}

The main components are:
\begin{itemize}
\item a graphical user interface (GUI);
\item an engine for processing user queries;
\item a page loader and (page to program) translator; and
\item a store of LogicWeb programs.
\end{itemize} 
The Prolog engine consists of three components: the query processing
engine, the program store, and the page loader and translator. The query
processing engine consists of the LogicWeb program interpreter (see \S8.1)
and predicates to translate user actions into the evaluation of specific
goals in LogicWeb programs. Downloaded LogicWeb programs are stored
inside the program store (actually as facts in a SWI-Prolog database).
The page loader and translator consist of predicates which download and
parse HTML documents to extract the clauses making up LogicWeb programs.

The system converts a user action (i.e., link selection or form
submission) into a query. It computes the result of the query with respect
to a program by invoking the LogicWeb interpreter. Processing of the
query may result in other pages being downloaded and translated into
LogicWeb programs. When query processing ends, the system formats the
result and shows it to the user via the browser.

The above architecture has been implemented using the NCSA Mosaic
browser and SWI-Prolog (for the Prolog engine). The two are integrated
with Mosaic's Common Client Interface (CCI) API~\cite{ld98b}.
This implementation is extended in \S8.

\section{Overview of the Security Model}

The LogicWeb security model assumes the following about the 
downloading and execution of programs:
\begin{itemize}
\item Program {\em source} is downloaded. Since the local host receives
the source rather than a compiled form (e.g., binaries or bytecode),
each goal is visible at the interpreter level. Also, if the source is
unencrypted then it is open to tampering -- it may be intercepted during
transmission and modified.

\item Downloaded programs are executed by an interpreter based on 
the operational semantics presented earlier. Such an interpreter
is described in~\cite{ld98b}, and is extended in \S8.

\item A LogicWeb program can invoke goals in other programs. 
Therefore, the model must handle communication between programs 
possessing different levels of trust. A less trusted 
program should not be able to use the privileges of a more trusted 
program.
\end{itemize} 

We have adopted a {\em sandbox} model, where each program is executed
in a controlled space, limiting its access to resources and
controlling the resources used. A sandbox for a LogicWeb program consists
of its interpreter and a {\em security policy}, or {\em policy} for short,
which controls the program's utilisation of system resources. 
A less trusted
LogicWeb program is assigned a policy which restricts resource usage,
while a more trusted program is granted a more liberal policy.
Security policies are defined by each user of the LogicWeb system.
 
A LogicWeb program accesses the local environment and system resources via
{\em system calls}, which are SWI-Prolog and LogicWeb built-in predicates.
LogicWeb built-ins include predicates for displaying pages, constructing
and parsing HTML documents, fast string matching, checking if a particular
program exists in the program store, and deleting programs from the
program store.

A policy specifies what system calls are permitted, and the ways they
can be used. This information is encoded as a set of predicates in a
{\em policy program} created by the LogicWeb system user. 
A policy program also states what programs can be
utilised via context switching. Policy programs may be stored at remote
sites or locally as long as they are protected from tampering. They can
be integrated into the system using the same downloading mechanism as
the rest of the LogicWeb system, including the use of composition to
combine policies.

\begin{sloppypar}
Assignment of policies to LogicWeb programs is {\em owner-based}.
The choice of policy program assigned to a LogicWeb program is determined
by how much the LogicWeb program is trusted which, in turn, depends on
the owner of the program. The owner is identified 
by a mechanism external to the LogicWeb language, by 
authenticating the LogicWeb program using a PGP (Pret\-ty Good Privacy) 
digital signature~\cite{garfinkel95}.

\end{sloppypar}

\begin{sloppypar} 
Policy programs restrict the use of system calls, thereby preventing
integrity and privacy attacks. Denial-of-service attacks are addressed
by controlling the use of resources using policy programs (as shown in
\S9.1) and meta-interpreters (discussed in \S9.2).  
\end{sloppypar}

\section{Digital Signatures for LogicWeb Programs}

Information can be encrypted in PGP using a secret {\em private key}, and
decrypted using a corresponding {\em public key} which is distributed
publically. PGP keys are used for {\em authentication} in the following way.
Suppose A encrypts a message and sends it to B. B wants to ensure that
the encrypted message is really from A, and that the message has not been
tampered with. To do this, B attempts to decrypt the message using A's
public key. If the decryption succeeds, it means that the message was
encrypted using A's private key and that it has not been modified. Also,
since the key is only known to A, the message must have come from A.
If the decryption fails, B cannot conclude that A encrypted the message.

PGP keys can also be used to {\em sign} documents or programs. Figure~3 shows
the use of a digital signature when downloading a program from A to B,
and the process of assigning a policy program. A PGP digital
signature is created by first mapping the program
to a string (actually, a single large number) using the MD5
algorithm~\cite{garfinkel95}, which almost uniquely identifies
the message.\footnote{In theory, two different messages could
map to the same MD5 string, although this has never been known to
occur~\cite{garfinkel95}.} The MD5 string is then encrypted using
the sender's private key, and this encrypted MD5 string becomes the
digital signature. The receiver of a signed program (the program and its
signature) uses the sender's public key to decrypt the
signature, producing the original MD5 string and the PGP ID. 
The receiver then
checks that the program corresponds to the MD5 string, i.e. the program
has not been modified in any way, and assigns a policy to the
program by consulting a database maintained in the
system which maps PGP IDs to policy program IDs.
Note that the system uses three kinds of IDs: PGP IDs, policy program IDs and
IDs of an application's LogicWeb programs, and that  
since policy programs are LogicWeb programs, policy program IDs are LogicWeb
program IDs. Unsigned programs,
or programs where authentication failed, should not be trusted, and
are given restricted access to system resources. The system stores the
downloaded program, and records the policy assignment for later use in query
evaluation. Policy programs not already present in the system
are downloaded during query evaluation. \S8 provides details on policy
assignment and use.

\begin{sidewaysfigure}
\setlength{\epsfxsize}{7.0in} 
\begin{center}
\boxps{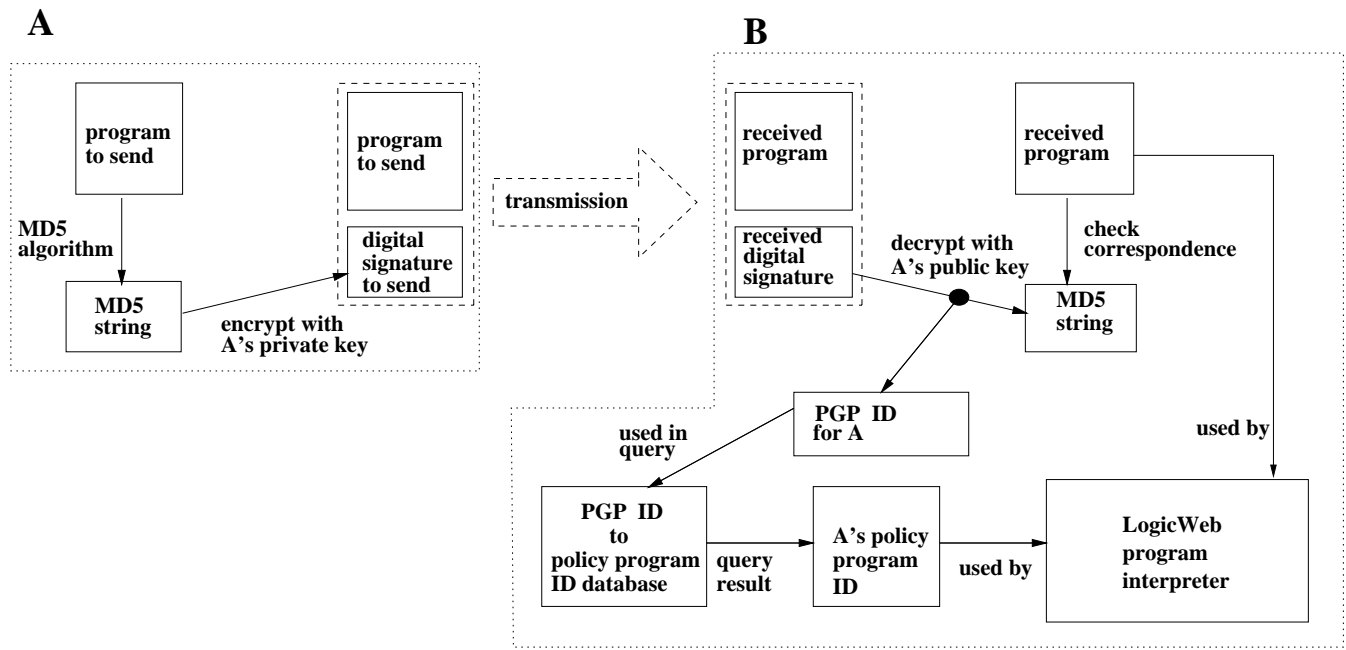}
\end{center}
\caption{Downloading a program from A to B,
and the subsequent assignment of a policy program.}
\end{sidewaysfigure}

A digital signature proves who sent the program, and that the program
was not altered either by error or design. The signature also provides
non-repudiation, which means that the sender cannot easily disavow his
signature on the program.

One reason for using an MD5 string is that it is faster to encrypt
and decrypt than the entire program. Also, since the program is not
encrypted, the program source is available even if authentication fails,
and so can still be executed (though with limited privileges). However,
if program confidentiality is required during transmission, the code
can be encrypted.

We utilise PGP since it is difficult to break, equipped
with public and private keys generation, widely available, supported
on a range of operating system platforms, and one of the most popular
encryption techniques. PGP public keys can be obtained either through
personal communication between the LogicWeb system user and program
owners, or via the Web: PGP public keys are increasingly being
placed on homepages, where they can be readily retrieved by the LogicWeb
system. The distribution of PGP keys is discussed further in \S11.

\section{Specifying Security Policies}

\begin{sloppypar}
A LogicWeb program accesses system resources via system
calls {\em and} context switching. The resources used via 
context switching are socket connections for downloading pages 
and local storage for programs. 

A policy program defines three predicates for controlling
the system calls and context switching of its designated LogicWeb 
programs. The predicates are:
\begin{itemize}
\item \texttt{valid\_program(Type, URL)}: this predicate
specifies the LogicWeb programs which can be used
via context switching. For example, restricting 
programs to those from the domain \texttt{www.cs.mu.oz.au} is 
given by:
\begin{alltts}
valid_program(get, URL) :- contains(URL, "http://www.cs.mu.oz.au/").
\end{alltts}
The following query will fail when using that policy 
since the goal refers to an invalid domain:
\begin{alltts}
?- lw(get, "http://www.cs.rmit.edu.au/")#>h_text(Src).
\end{alltts}
 
\item \texttt{valid\_systemCall(Call)}: this predicate defines
the set of system calls a program is allowed to invoke. 
\texttt{Call} is a term representing the form the system call may
take.

\item \texttt{call\_system(Call)}: this predicate is a wrapper for
the allowed predicates defined by \texttt{valid\_systemCall/1}. 
\texttt{Call} is a term representing the goal which is to be 
invoked in a special way. Typically, \texttt{call\_system/1} is
used to implement more restrictive versions of system calls. 
\end{itemize}

The difference between
\texttt{valid\_systemCall/1} and \texttt{call\_system/1} is
{\em execution}. Call patterns specified in \texttt{valid\_systemCall/1}
are used to {\em test} system calls, rejecting ones which do not
match the necessary requirements. \texttt{call\_system/1} is used
to {\em execute} system calls in novel (usually restricted)
ways. A policy program could be written with
the body code of \texttt{valid\_systemCall/1} implemented in
\texttt{call\_system/1}, doing away with the need for the
\texttt{valid\_systemCall/1} predicate.
However, the distinction between testing and execution would 
then be much less clear.

\end{sloppypar}

\begin{sloppypar}
The following policy program permits
the retrieval (using the GET method) of all URLs
except \texttt{http://www.cs.mu.oz.au/\~{}swloke/private.html}.
It disallows all calls to \texttt{system/1} and
\texttt{open/3}, except when \texttt{open/3} reads \texttt{dump.txt}:
\end{sloppypar}
\begin{alltts}
valid_program(get, URL) :-
  URL \(\backslash\)== "http://www.cs.mu.oz.au/~swloke/private.html".

valid_systemCall(open('/home/pgrad/swloke/lws/dump.txt', read, S)).
valid_systemCall(Call) :-
  Call \(\backslash\)= open(_, _, _),
  Call \(\backslash\)= system(_).

call_system(Call) :- 
  built_ins:call_builtin(Call).
\end{alltts}
\texttt{call\_builtin/1} is a system predicate which carries out
type checking on a call before executing it. For example, the 
\texttt{call\_builtin/1} clause for \texttt{open/3} is:
\begin{alltts}
call_builtin(open(FileName, Mode, Stream)) :-
  atom(FileName),
  ( Mode = read
  ; Mode = write
  ),
  var(Stream),
  open(FileName, Mode, Stream).
\end{alltts}
\texttt{FileName} must be an atom, \texttt{Mode} either \texttt{read}
or \texttt{write}, and \texttt{Stream} a variable. Type checking 
increases the robustness of the system by preventing instantiation faults 
when the arguments are the wrong type.

\medskip

We briefly consider two examples that show the flexibility of 
security policies.

\subsection{A Historical Policy}
State information can be used to implement 
a history-based policy. For example, the following policy lets
a program carry out context switching (i.e.\ execute goals
such as \texttt{lw(get, URL)\#>Goal}) until a file is accessed.
A file access is detected by having \texttt{call\_system/1} monitor
\texttt{open/3} calls.
\begin{alltts}
file_accessed(no).               % state information: no file has been accessed

valid_program(_, _) :-           % only context switch when no
  file_accessed(no).             % files have been accessed

call_system(Call) :-             % process non-open/3 calls
  Call \(\backslash\)= open(_, _, _),
  built_ins:call_builtin(Call).
call_system(open(F, R, S)) :-    % process open/3 call
  built_ins:call_builtin(open(F, R, S)),
  (file_accessed(no) ->
    retract(file_accessed(no)),  % record file access
    assert(file_accessed(yes))
  ;
    true
  ).
\end{alltts} 
Once \texttt{open/3} has been called, \texttt{file\_accessed/1} will
be changed to hold the value \texttt{yes}. This will cause
subsequent calls to \texttt{valid\_program/2} to fail, disabling
context switching for the program.

\subsection{Levels of Trust}

It is straightforward to build policies with varying levels of trust 
for different programs. In the following example we use three levels:
dangerous, ok, and safe. A `dangerous' program cannot download
any programs, or execute any system calls.
An `ok' program can download programs but it can only write
to the directory \texttt{/tmp} and cannot use \texttt{system/1}
to delete files. A `safe' program has no restrictions placed upon it,
except that a message is printed out when a file deletion succeeds.
Each level of trust is represented by a separate policy program.

The policy program for `dangerous':
\begin{alltts}
valid_program(_, _) :- fail.

valid_systemCall(_) :- fail.

call_system(_) :- fail.
\end{alltts}
Any call that uses system resources will fail.

The policy program for `ok' is given as:
\begin{alltts}
valid_program(_, _).                  % no restrictions on downloads

valid_systemCall(open(Path, write, _)) :-
  !, append("/tmp/", _, Path).        % only write to /tmp
valid_systemCall(system(Cmd)) :-
  append("rm ", _, Cmd), !, fail.     % disallow file deletions
valid_systemCall(_).                  % everything else allowed
  
call_system(Call) :-                  % no restrictions on execution
  built_ins:call_builtin(Call).
\end{alltts}

The policy program for `safe':
\begin{alltts}
valid_program(_, _).                 % no restrictions on downloads

valid_systemCall(_).                 % all system calls permitted

call_system(Cmd) :- 
  built_ins:call_builtin(Cmd),
  (append("rm ", _, Cmd) ->          % report deletions when executed
     report_deletion(Cmd)
  ;
     true
  ).
\end{alltts}
We could call \texttt{report\_deletion/1} in \texttt{valid\_systemCall/1}
but reporting a file deletion has nothing to do with testing the
validity of a system call; it is a diagnostic associated with
execution.

\section{Combining Security Policies}

\begin{sloppypar}
A LogicWeb program must not be allowed to perform an illegal system
operation by invoking a goal in a more privileged program. Hence,
the security model must ensure the following: 
\begin{enumerate} 
\item {\em Context switching must not transfer privileges between
programs.} The system has to ensure that an untrusted program does not
illegally access resources by invoking a goal in a trusted program. For
example, assume that a program \texttt{P} is not allowed to read a file
(as determined by its policy) while another program \texttt{Q}
can (as determined by \texttt{Q}'s policy). This means that \texttt{P}
should not be able to call a goal like \texttt{Q\#>read\_file(Contents)}
to read a file. Clearly, it is too simplistic to validate a call to
\texttt{read\_file/1} in \texttt{Q} using only the policy of \texttt{Q};
the policy of \texttt{P} must also be considered.

In a similar manner, a LogicWeb goal cannot be allowed to perform an
illegal system operation because it is invoked in a trusted program.
For example, if \texttt{A} can read a file which \texttt{B} can not,
then the goal \texttt{B\#>read\_file(Contents)} invoked in \texttt{A}
should fail. This means that although \texttt{A} uses \texttt{B},
\texttt{A} should not pass its privileges to \texttt{B}, since the
system trusts \texttt{A} but not \texttt{B}. Hence, the policies of both
\texttt{A} and \texttt{B} must be consulted to determine if a
\texttt{read\_file/1} call should be allowed, as Figure~4 illustrates.

\begin{figure*}
\setlength{\epsfxsize}{4in} 
\begin{center}
\boxps{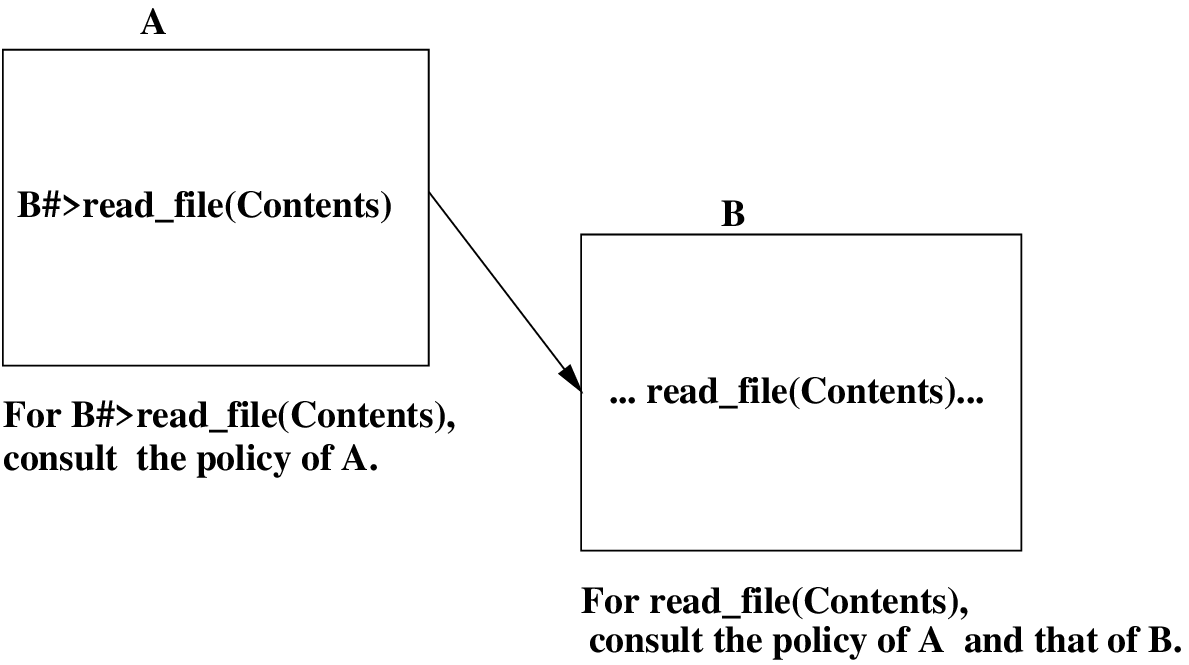}
\end{center}
\caption{ The invocation of \texttt{B\#>read\_file(Contents)}
in \texttt{A}. }
\end{figure*}

When more than two programs are involved, a similar but slightly
more complex situation arises. For example, consider a goal 
\texttt{B\#>(C\#>G)} invoked in the program \texttt{A}, where 
\texttt{G} is a system call. If \texttt{G} is allowed by the policies
of \texttt{A} and \texttt{C}, but \texttt{B} does 
not permit \texttt{G}, then \texttt{B} is performing an illegal 
system operation through \texttt{C}.

Essentially, disregarding any one of the policies potentially allows an
illegal system operation. This means that to determine if \texttt{G}
should be allowed, {\em all} the policy programs used by the code must
be considered.

Figure~5 shows the changes in goal evaluation contexts starting from the
goal \texttt{B\#>(C\#>G)} in program \texttt{A} and ending with goal
\texttt{G} in \texttt{C}. The policies of all the programs involved
in the evaluation of the goal starting from the first program (i.e.,
\texttt{A}) are required. Hence, the proof rules for the security model
must incorporate a notion of the {\em current} set of policy programs
which grows as goal evaluation progresses.

\begin{figure*}
\setlength{\epsfxsize}{4.4in} 
\begin{center}
\boxps{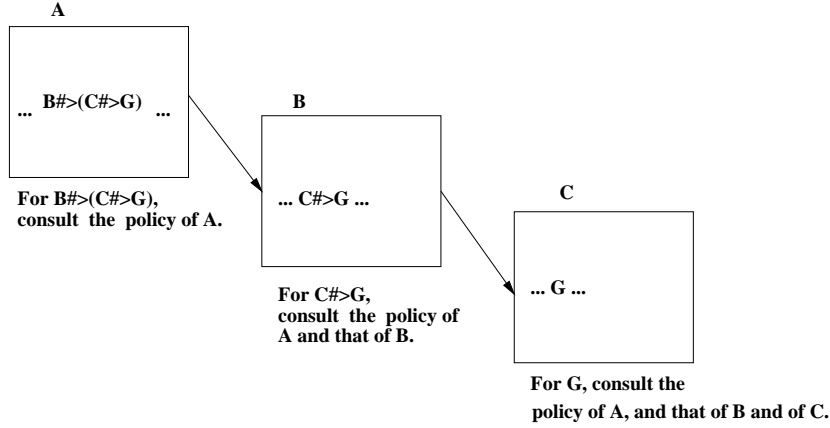}
\end{center}
\caption{The evaluation \texttt{B\#>(C\#>G)} in program \texttt{A}.}
\end{figure*}

\item {\em LW-compositions must not transfer privileges between
programs.} A LogicWeb program can invoke the clauses of another 
program in a LW-composition. For example, consider 
a goal \texttt{G} evaluated in the LW-composition of programs
\texttt{S} and \texttt{T}: \texttt{(S + T)\#>G}. \texttt{G}'s 
evaluation may utilise some clauses from \texttt{S} and some from
\texttt{T}, so in general \texttt{G} must be allowed by {\em both}
the policies for \texttt{S} and \texttt{T}.
\end{enumerate}
\end{sloppypar}

\begin{sloppypar}
The observations made above can be made more formal by restating them
using the LW-encapsulation and LW-intersection operators 
(\texttt{@} and \texttt{*}). Namely, given a set of policies
(e.g., \{\texttt{P},\texttt{Q}\}), the allowable set of system calls
and downloads are those permitted by the composition
\texttt{@P} \texttt{*} \texttt{@Q}. 

LW-encapsulation is used to express that each policy must 
separately validate the system calls and downloads. 
LW-intersection is used to represent that a given
system call or download must be permitted by {\em every}
policy program. For example, an \texttt{open/3} call is permitted by 
both \texttt{P} and \texttt{Q} when the following goal succeeds: 
\begin{alltts}
?- (@P * @Q)#>valid_systemCall(open(_,_,_)).
\end{alltts}
\end{sloppypar}

\section{Enforcing Security Policies}

This section defines a new derivation relation, extending 
the operational semantics in \S2.2.3 to use policy programs.
Previously, a goal was evaluated in the current context, with 
the current set of created programs. In the new derivation
relation, we also consider the current set of 
policy programs. The derivation relation is extended as follows: 
\begin{quote}
For any goal formula $G$ and program expression $E$, 
 \[ \Sigma, S, E \vdash^{S'}_{\theta} G \] 
denotes the fact that there exists a {\em top-down derivation}
of $G$ in $E$ starting with the set $S$ of existing 
LogicWeb programs and the {\em ordered} set of policy programs $\Sigma$,
and ending with the computed answer substitution $\theta$ and 
created program set $S'$.
\end{quote}

The proof rules which define this new relation are an extension 
of the rules in \S2.2.3. Rules (1) to (8) are modified to form 
corresponding rules for the new derivation relation by using 
the following syntactic mapping: 
\begin{quote}
Replace every occurrence of an expression involving the derivation 
relation:
 \[S, E \vdash^{S'}_{\theta} G\] with the corresponding expression:
 \[\Sigma, S, E \vdash^{S'}_{\theta} G\]
\end{quote}
For instance, the following
rule is obtained by applying this mapping to rule (2) (conjunction):
\begin{equation*}
    \frac{\Sigma, S, E \vdash_{\theta}^{S'} G_1\;\;\;\wedge\;\;\;\Sigma, S', E \vdash_{\gamma}^{S''} G_2\theta}{\Sigma, S, E \vdash_{\theta \gamma}^{S''} G_1,G_2}
\end{equation*}
The same set of policy programs $\Sigma$ is used for each of the
conjuncts since they occur in the same rule (and hence, in the same 
program).

\subsection{Mapping from Programs to Policies}

To aid the discussion which follows, two functions are defined: one maps
a LogicWeb program to its policy program, and the other relates a set
of LogicWeb programs to their policy programs. $\Phi$ denotes the set
of all policy programs used by the system.
 
A policy program is permitted free access to system resources, and 
is not itself assigned a policy program. 

\begin{Def}[policy program assignment]
The function \[pol: (LWProgramIDs~\backslash~ids(\Phi)) \rightarrow \Phi \]
takes a program identifier (which is not that of a policy program)
and returns a policy program for that program from $\Phi$.
\end{Def}
$pol$ is not defined on policy programs, 
and the empty program $\xi$ is not assigned a policy program.

\begin{Def}[policies for a set of programs]
The function \[pols: \wp (LWProgramIDs~\backslash~ids(\Phi)) \rightarrow \wp (\Phi) \]
takes a set of (non-policy) program identifiers $I$ and returns the
set of policy programs assigned to the programs in $I$. 
$pols$ is defined by:
\begin{align*}
pols(I) =~& \{pol(i)~|~i \in I\}
\end{align*}
\end{Def}

\subsection{Context Switching Revisited}

$\texttt{*}\Sigma^{\texttt{@}}$ denotes the LW-intersection 
of the LW-encapsulation of the policy programs in $\Sigma$ (i.e., 
if $\Sigma=\{P_1, ..., P_n\}$ then $\texttt{*}\Sigma^{\texttt{@}}=\texttt{@}P_1~\texttt{*}...\texttt{*}~\texttt{@}P_n$),
and $G$ denotes a goal.
Evaluating a goal against $\texttt{*}\Sigma^{\texttt{@}}$ has the effect
of evaluating the goal against each program in $\Sigma$ separately.

\begin{sloppypar}
The new definition of the context switching rule (9)  is:
\begin{align*}
& \;\;\Sigma \not= \emptyset\;\;\wedge\\
& \;\;\biggl\{\;Where~each~\texttt{lw(Type$i$, URL$i$)} \in expids(F')~\backslash~ids(\Phi),\\
& \;\;\;\;\;\;\;\;\;\;\;\emptyset, S, \xi \vdash^{S'}_{\gamma} \texttt{*}\Sigma^{\texttt{@}}\texttt{\#>}\texttt{(valid\_program(Type1, URL1)$,$} \\
& \;\;\;\;\;\;\;\;\;\;\;\;\;\;\;\;\;\;\;\;\;\;\;\;\;\;\;\;\;\;\;\;\;\;\;\;\;\;\texttt{$\ldots,$valid\_program(TypeN, URLN))} \; \biggr\} \;\;\wedge \\
&\frac{\;\;\;I \subseteq ids(S')\;\;\wedge\;\;\;(pols(I~\backslash~ids(\Phi)) \cup \Sigma), S', F' \vdash_{\theta}^{S''} G\;\;\;\;\;\;\;\;\;\;\;\;\;\;\;\quad\quad\quad}
{\;\;\;\Sigma, S, E \vdash_{\theta}^{S''} F~\texttt{\#>}~G} \tag{9'a}
\end{align*}
The new rule has been augmented with:
\begin{enumerate}[i.]
\item A test to determine if each non-policy program 
to be used in the new context is allowed by the
current set of policy programs $\Sigma$.  
The test is done by invoking the predicate
\texttt{valid\_program/2} in $\texttt{*}\Sigma^{\texttt{@}}$ for each such
non-policy program identified by \texttt{lw(Type$i$, URL$i$)} 
using a goal of the form:
\begin{align*}
 \emptyset, S, \xi \vdash^{S'}_{\gamma} \texttt{*}\Sigma^{\texttt{@}} \texttt{\#>} & \texttt{(valid\_program(Type1, URL1)$,\ldots,$}\\
& \texttt{valid\_program(TypeN, URLN))}
\end{align*}
\texttt{N} is the number of program identifiers in 
$expids(F')~\backslash~ids(\Phi)$. 
A LogicWeb goal is used when evaluating the \texttt{valid\_program/2} goals in
order to download the policy programs.
Note that goal evaluation fails if not all the policy programs 
are downloaded. Also, a policy program can invoke non-policy programs 
in its rules, and the evaluation of the \texttt{valid\_program/2} goals 
begins with an empty policy program set. The rule for the case where 
$\Sigma$ is empty is given below.

Note that we test all the programs
in $F' = insertCC(F, E)$ instead of $F$, which means that we test the
programs in the current context as well. 

\item An extension of $\Sigma$ with new policy programs.
New policy programs are added to
the front (left) of the ordered set $\Sigma$: 
\[pols(I~\backslash~ids(\Phi)) \cup \Sigma\]
The last (rightmost) program in the ordered set is, chronologically,
the first policy program, and its significance is explained below.
\end{enumerate}
\end{sloppypar}

The following variant of the context switching rule caters for 
when $\Sigma$ is empty. In that case,
no policy programs are employed, no checks are made:
\begin{align*}
&\frac{\;\;\;I \subseteq ids(S')\;\;\wedge\;\;\;pols(I~\backslash~ids(\Phi)), S', F' \vdash_{\theta}^{S''} G\;\;\;\;\;\;\;\;\;\;\;\;\;}
{\;\;\;\emptyset, S, E \vdash_{\theta}^{S''} F~\texttt{\#>}~G} \tag{9'b}
\end{align*}

\medskip

In (i), we applied \texttt{valid\_program/2} to all the 
programs used in the goal's program expression \emph{and}
all the programs in the current context. We illustrate
why by considering an example:
\begin{alltts}
valid_program(get, 'A').
valid_program(get, 'B').
valid_program(get, 'C') :- fail.   % may be omitted but added for clarity
\end{alltts}
If the goal \texttt{((A+B)*C)\#>G} is used in \texttt{A} then
we would expect it to fail since \texttt{A} is not permitted to use 
\texttt{B} in context switching. Suppose instead that the goal
was \texttt{((\#)*C)\#>G} and the current context (as represnted
by \texttt{\#}) was \texttt{A+B}. The goal is equivalent to
the earlier one; should it be permitted or not?

If the goal is analysed by testing only \texttt{C} and ignoring the
current context, then \texttt{A} would be allowed to use
\texttt{B} in context switching, so violating \texttt{A}'s policy.
However, this behaviour appears harmless since \texttt{B} must
already have been retrieved if it is in the current context,
and so its utilisation will require no additional resources.

This would not be the case if the semantics of 
\emph{add\_programs} in Definition 2.2 were different. Suppose
that \texttt{A}'s use of \texttt{B} did result in the retrieval
of a new version of \texttt{B}, then resources would be used.
Or suppose that programs are only retrieved when their clauses
are actually needed, and so \texttt{B} may not have been
retrieved at the time of \texttt{G}'s evaluation.

By conservatively testing the current context, we stay
faithful to \texttt{A}'s policy: \texttt{A} cannot use
\texttt{B} via context switching, regardless of whether
\texttt{B} is downloaded. This design is more general
since it makes the security model more tolerant
of changes in the system's download and storage mechanisms.
Also, the intention of \texttt{A}'s policy is arguably
to unconditionally bar \texttt{A} from using \texttt{B}
via context switching.

We could easily implement the less conservative view:
namely that \texttt{A} cannot use \texttt{B} via context switching,
unless \texttt{B} is already downloaded (or in the current
context). \texttt{A}'s policy program would become:
\begin{alltts}
valid_program(get, 'A').
valid_program(get, 'B').
valid_program(get, 'C') :- program_exists('B').
\end{alltts}
\texttt{program\_exists/1} is a LogicWeb built-in which 
succeeds if the specified program is in the local cache.

\subsection{System Call Rules}

Two new rules called {\em system call rules} are added
to specify how policy programs are utilised with system calls. 
The first rule specifies a test to determine if a goal
is a valid system call. It invokes \texttt{call\_system/1}
in the last policy program in $\Sigma$, which is the policy
for the main program of the application. This means 
that any state changes carried out by \texttt{call\_system/1}
(for instance, see the example in \S 5.1) will be done in
the main program's policy program.
\begin{align*}
& \;\;\Sigma \not= \emptyset\;\;\wedge\\
& \;\;\;\emptyset , S,  \xi \vdash^{S'}_{\gamma} \texttt{*}\Sigma^{\texttt{@}}\texttt{\#>}\texttt{valid\_systemCall($G$)}\;\;\wedge \\
& \;\;\;\Sigma=\Sigma' \cup P_n\;\;\wedge \\
& \frac{\;\;\;\emptyset, S', \xi  \vdash^{S''}_{\theta} P_n\texttt{\#>}\texttt{call\_system($G$)}\;\;\;\;\;\;\;\;\;\;\;\;\;\;\;\;\;\;\;\;\;\;\;}
{\;\;\;\Sigma, S, E \vdash^{S''}_{\theta} G}  \tag{10'a}
\end{align*}
Substitutions resulting from calling \texttt{valid\_systemCall/1} are discarded.
Instead, the substitutions computed from
\texttt{call\_system/1} are kept. As before, proofs occurring in 
policy programs proceed with an empty policy program set.

For the case where $\Sigma = \emptyset$, the rule is
{\footnotesize
\begin{align*}
 \frac{\;\;\;\emptyset, S, E  \vdash^{S'}_{\gamma} \texttt{built\_ins:builtin($G$)}\;\;\;\wedge\;\;\;\emptyset, S', E  \vdash^{S''}_{\theta} \texttt{built\_ins:call\_builtin($G$)}}
{\;\;\;\emptyset, S, E \vdash^{S''}_{\theta} G} \tag{10'b}
\end{align*}
}

A call to a built-in predicate in the module \texttt{built\_ins} is
represented by:
\begin{equation*}
\frac{G~succeeds~with~\theta}{\emptyset, S, E \vdash^{S}_{\theta} \texttt{built\_ins:($G$)}} \tag{11}
\end{equation*}
The evaluation of the goal is outside 
the scope of the inference rules, but $\theta$ is assumed to 
be the computed answer substitution.

\subsection{Properties of the Inference Rules}

We show how the inference rules disallow 
illegal system operations. First, {\em context sequence} and 
{\em illegal system operation} are defined.

Goal derivation involving LogicWeb goals may result in several 
sequences of context swit\-ches, each represented by a sequence of contexts.
For example, the sequence of contexts in Figure~5 is:
\[\texttt{A},\;\;\texttt{B},\;\;\texttt{C}\]
Note that the only operational semantic rules which change the context 
are the context switching rules (i.e. 9'a and 9'b). 

\begin{Def}[context sequence]
Given a sequence of applications of inference rules
in a top-down derivation, suppose that there are $n-1$ applications of 
the context switching rules. Further suppose that the $i$th application 
of these rules causes the context to switch from $E_i$ to $E_{i+1}$, 
where $1 \leq i \leq n-1$. The $context$ $sequence$ is
the sequence of contexts $E_1,\ldots,E_n$.
\end{Def}

Informally, given a context sequence $E_1,\ldots,E_n$,
an illegal system operation is performed in the context $E_i$ if 
a system call, or the use of a program in
context switching, is disallowed by the policy of some program in
$E_i$, but is attempted in $E_i$ or some later context. 

\begin{sloppypar}
\begin{Def}[illegal system operation]
Given a context sequence $E_1,\ldots,E_n$, for some
$E_i$, $P \in pols(expids(E_i)~\backslash~ids(\Phi))$, $i \leq j \leq n$, 
and system call $G$, an illegal system operation
is performed in $E_i$ if the goal \texttt{call\_system($G$)} is invoked 
when the context of $G$ is $E_j$ but 
$\emptyset, S, P \vdash^{S'}_{\theta} \texttt{valid\_systemCall($G$)}$ 
does not hold for any $S$, $S'$, and $\theta$,
or the oracle function is called on the identifier \texttt{lw(Type, URL)}
when the current context is $E_j$
but $\emptyset, S, P \vdash^{S'}_{\theta} \texttt{valid\_program(Type, URL)}$
does not hold for any $S$, $S'$, and $\theta$.
\end{Def}
\end{sloppypar}

The context switching rules guarantee that all policy programs which must
be consulted during a derivation are present.

\begin{Lem}
Given the context sequence $E_1,\ldots,E_n$ for a top-down derivation, 
for any $i \in \{1,\ldots,n\}$, at the node with context $E_i$,
the current set of policy programs
$\Sigma$ contains the policy programs of all the non-policy
programs occurring in $E_1,\ldots,E_i$. 
\end{Lem}
\begin{proof}
The proof is by induction on $i$.
For $i=1$, the goal evaluation starts in the empty program $\xi$ 
with $\Sigma=\emptyset$. 
For $i\geq 2$, (by the inductive hypothesis) when the goal is 
evaluated in $E_{i-1}$, $\Sigma$ contains the policy programs of 
all the non-policy programs occurring in
$E_1,\ldots,E_{i-1}$. $E_{i-1}$ changes to $E_i$ using one of 
the context switching rules. This rule updates $\Sigma$ (say, to $\Sigma'$) 
by adding to $\Sigma$ the policy programs of the non-policy programs
occurring in $E_i$.
Hence, the goal evaluation continues in $E_i$ with $\Sigma'$
containing all the policy programs for the non-policy
programs in $E_1, \ldots, E_i$.
\end{proof}

The theorem below implies soundness of the security model:
all goals which have a successful derivation have the desired
security property of not performing an illegal system operation.

\begin{sloppypar}
\begin{The}[safety property]
No illegal system operation is performed in any context during a 
top-down derivation using the above rules. 
\end{The}
\begin{proof}
Given the context sequence $E_1,\ldots,E_n$,
for any $i \in \{1,\ldots,n\}$,
suppose that an illegal system operation was performed in 
$E_i$, say when the current context was $E_j$, where $i \leq j \leq n$,
the set of policy programs was $\Sigma$, and the set of created programs 
was $S$. To perform this operation either (1) the system call
rule for $\Sigma \not= \emptyset$ or (2) the new context switching 
rule for $\Sigma \not= \emptyset$ must have been used. In case (1), 
according to the definition of the system call rule used, since 
\texttt{call\_system($G$)} was invoked, the goal:
\[\emptyset, S \cup \Sigma, 
\texttt{*}\Sigma^{\texttt{@}} \vdash^{S'}_{\gamma} 
\texttt{valid\_systemCall($G$)}\] 
must have succeeded. In case (2), suppose that the LogicWeb goal
was $F\texttt{\#>}G$ and $F' = insertCC(F, E_j)$, and 
the oracle function was called on the identifier 
$\texttt{lw(Type, URL)} \in expids(F')~\backslash~ids(\Phi)$, then 
according to the definition of the context switching rule used, 
the goal:
\[ 
\emptyset, S \cup \Sigma, \texttt{*}\Sigma^{\texttt{@}} \vdash^{S'}_{\gamma} 
\texttt{valid\_program(Type, URL)}\]
must have succeeded. But by the definition of an illegal system operation, 
for some $P \in pols(expids(E_i)~\backslash~ids(\Phi))$, in case (1), 
$\emptyset, T, P \vdash^{T'}_{\theta} \texttt{valid\_systemCall($G$)}$ 
does not hold for any $T$, $T'$, and $\theta$, and in case (2),
$\emptyset, T, P \vdash^{T'}_{\theta} \texttt{valid\_program(Type, URL)}$
does not hold for any $T$, $T'$, and $\theta$. In either case, it must 
have been that $P \not\in \Sigma$. By the above lemma, $\Sigma$ contains 
the policy programs of all the non-policy programs occurring in
$E_1,\ldots,E_j$, namely, 
$\Sigma \supseteq pols(expids(E_i)~\backslash~ids(\Phi))$. This means 
$P \in \Sigma$, and hence there is a contradiction. 
\end{proof}
\end{sloppypar}

\section{Implementation}

\subsection{Interpreter with Security Mechanisms} 

\begin{sloppypar}
The inference rules presented above provide the basis
for an interpreter for evaluating goals in the presence of policy programs.
Program~1 shows this interpreter, which is an extension of the
one described in~\cite{ld98b}. 

\end{sloppypar}
\begin{program}
\begin{center}
\begin{alltts} 
\def\baselinestretch{1.0} 
{\footnotesize
{\rm {\em \% demo/3 with LogicWeb goal}}
demo(PL, E, F#>G) :- 
  establish_context(PL, F, E, PL, NPL, F1), 
  demo(NPL, F1, G).
demo(PL, E, G) :- 
  allowed_systemCall(PL, G), invoke_systemCall(PL, G).
demo(_PL, P, built_ins:G) :- 
  pgpID_to_policyID(_, P), call(built_ins:G). 
demo(_PL, _E, true).
demo(PL, E, (A, B)) :- demo(PL, E, A), demo(PL, E, B).
demo(PL, E, A) :- select_clause(PL, E, (A :- B)), demo(PL, E, B).

{\rm {\em \% establish a context}}
establish_context(OPL, E + F, C, PL, PL2, E1 + F1) :-
  establish_context(OPL, E, C, PL, PL1, E1), 
  establish_context(OPL, F, C, PL1, PL2, F1).
establish_context(OPL, E * F, C, PL, PL2, E1 * F1) :-
  establish_context(OPL, E, C, PL, PL1, E1), 
  establish_context(OPL, F, C, PL1, PL2, F1).
establish_context(OPL, E / P, C, PL, PL2, E1 / P1) :-
  establish_context(OPL, E, C, PL, PL1, E1), 
  establish_context(OPL, P, C, PL1, PL2, P1).
establish_context(OPL, @E, C, PL, PL1, @E1) :- 
  establish_context(OPL, E, C, PL, PL1, E1).
establish_context(OPL, (/)<>(E, L), C, PL, PL2, (/)<>(E1, L1)) :-
  establish_context(OPL, E, C, PL, PL1, E1), 
  establish_contextL(OPL, L, C, PL1, PL2, L1).
establish_context(OPL, Op<>L, C, PL, PL1, Op<>L1) :- 
  establish_contextL(OPL, L, C, PL, PL1, L1).
establish_context(OPL, lw(T, U), _C, PL, NPL, lw(T, U)):-
  allowed_program(OPL, lw(T, U)), download(T, U), 
  add_policyID(lw(T, U), PL, NPL).
establish_context(OPL, (#), C, PL, PL, C) :-
  allowed_programs(OPL, C).
 
establish_contextL(_OPL, [], _C, PL, PL, []).
establish_contextL(OPL, [E|Es], C, PL, PL2, [E1|Es1]) :-
  establish_context(OPL, E, C, PL, PL1, E1), 
  establish_contextL(OPL, Es, C, PL1, PL2, Es1).  

{\rm {\em % predicate to ensure that only allowed system calls are invoked}}
allowed_systemCall([], G) :- 
  built_ins:builtin(G).
allowed_systemCall([Pol|Pols], G) :-
  demo([], empty, Pol#>valid_systemCall(G)), 
  allowed_systemCall(Pols, G).
}
\end{alltts}

\caption{The interpreter for pure LogicWeb programs which use 
 policy programs.}
\end{center}
\end{program}

\addtocounter{program}{-1}
\begin{program}
\begin{center}
\begin{alltts} 
\def\baselinestretch{1.0} 
{\footnotesize
{\rm {\em % predicate to invoke system calls}}
invoke_systemCall([], G) :- 
  built_ins:call_builtin(G).
invoke_systemCall(PL, G) :-
  last(P, PL), demo([], empty, P#>call_system(G)).

{\rm {\em % add a policy program ID to list of IDs}}
add_policyID(Id, PL, PL) :- 
  pgpID_to_policyID(_, Id), !.% no policy program for policy programs
add_policyID(Id, PL, NPL) :-
  policyID(Id, NewPolicyId),
  (member(NewPolicyId, PL) ->
     NPL = PL
  ;
     NPL = [NewPolicyId|PL]
  ).

{\rm {\em % predicate to ensure that only allowed programs are downloaded}}
allowed_program([], _Id).
allowed_program([Pol|Pols], lw(Type, URL)) :-
  demo([], empty, Pol#>valid_program(Type, URL)), 
  allowed_program(Pols, lw(Type, URL)).  

{\rm {\em \% select a clause from the program store}}
select_clause(_PL, lw(Type, URL), A :- B) :- 
  lw(Type, URL)::(A :- B).
select_clause(PL, E + _F, A :- B) :- 
  select_clause(PL, E, A :- B).
select_clause(PL, _E + F, A :- B) :- 
  select_clause(PL, F, A :- B).
select_clause(PL, E * F, A :- (B,C)) :- 
  select_clause(PL, E, A :- B), select_clause(PL, F, A :- C).
select_clause(PL, E / P, A :- B) :- 
  select_clause(PL, E, A :- B), not defined(A, P).
select_clause(PL, @E, A :- true) :- demo(PL, E, A).

select_clause(PL, (/)<>(E, []), A :- B) :- 
  select_clause(PL, E, A :- B).
select_clause(PL, (/)<>(E, [P|Ps]), A :- B) :- 
  select_clause(PL, (/)<>[(E / P)|Ps], A :- B).

select_clause(PL, _Op<>[E], A :- B) :- 
  select_clause(PL, E, A :- B).
select_clause(PL, Op<>[E1,E2|Es], A :- B) :- 
  C =.. [Op, E1, E2], select_clause(PL, Op<>[C|Es], A :- B).

defined(A, P) :-
  functor(A, Functor, Arity), functor(H, Functor, Arity), 
  P::(H :- _B). 
}
\end{alltts}

\caption{(Continued)}
\end{center}
\end{program}

\texttt{demo/3} is derived from a standard vanilla Prolog
meta-interpreter~\cite{sb89,ss94}, and executes a goal (the third
argument of \texttt{demo/3}) against its program context (the second
argument). The first argument of \texttt{demo/3} stores the IDs of
downloaded policy programs. 

In the first clause of \texttt{demo/3}, \texttt{establish\_context/6}
returns the new policy program identifiers for the programs in 
the new context. 

\begin{sloppypar}
The second clause checks for and invokes
an allowed system call. In \texttt{allowed\_systemCall/2},
\texttt{valid\_systemCall/1} is invoked in each policy program.
Policy programs known to the system are recorded in
\texttt{pgpID\_to\_policyID/2}, which is described below.
\texttt{invoke\_systemCall/2} invokes the system call in
the last program mentioned in the list of
policy program identifiers (\texttt{PL}). This is the policy
program for the main program, as mentioned earlier.
\end{sloppypar}

The third clause of \texttt{demo/3} allows policy programs to
invoke goals in the Prolog module \texttt{built\_ins}.
\begin{sloppypar}
The last three clauses of \texttt{demo/3} implement the basic 
meta-interpreter. \texttt{select\_clause/3} in the final clause
evaluates the goal \texttt{A} by looking for
a suitable clause (\texttt{A :- B}) from the programs defined
by the program expression \texttt{E}. The clauses of downloaded
programs are held in the program store (see Figure~2) in the format
\texttt{lw(Type,URL)::(Clause)}.
\end{sloppypar}
\begin{sloppypar}
The interpreter is called using a goal such as:
\begin{alltts}
?- demo([], empty, lw(get, "http://main.program")#>query).
\end{alltts}
Goal derivation begins with the empty context without any policy programs
and a goal where \texttt{lw(get, "http://main.program")} is the main program of
an application.
\end{sloppypar}

\texttt{establish\_context/6} carries out several tasks. One of 
these is to expand the program expression \texttt{F} (which was part
of a context switching goal $F\texttt{\#>}G$), so that any
context operators (\texttt{\#}) in \texttt{F} are replaced by
the current context. This results in a new program expression 
\texttt{F1}. Note that
two copies of the original list of policy programs is passed into
\texttt{establish\_context/6} initially. 
The fourth and fifth arguments of
\texttt{establish\_context/6} represent the current extended
list and the newly extended list respectively. Because the current extended 
list may not be the original and the security tests 
are done using the original, the extra copy is supplied.

The last two clauses of \texttt{establish\_context/6} relate to the 
security model. The penultimate clause checks a new program ID
against the existing policies, and may download a new policy for
that program, which causes the list of policies IDs (\texttt{PL})
to be extended. 

\texttt{allowed\_program/2} checks a program ID (\texttt{lw(T,U)})
against the existing policies by calling \texttt{valid\_program/2}
against each policy in turn. This mimics the 
LW-intersection of LW-encapsulated policies described earlier.

\texttt{add\_policyID/3} attempts to retrieve a policy ID for the
specified program ID (\texttt{lw(T,U)}), and add it to the policy
IDs list. This process is complicated by the possibility that the program
ID may actually be a policy ID, which is ascertained by checking it against
\texttt{pgpID\_to\_policyID/2}. However, if \texttt{lw(T,U)} is an
ordinary LogicWeb program ID, then \texttt{policyID/2} will contain
a mapping from the program ID to its matching policy ID. How this
is achieved will be explained shortly.

The final clause of \texttt{establish\_context/6} checks all 
the programs in the current context against the existing 
policies. This check is required only once, but for simplicity, 
 we have permitted the code to perform
this check every time a context operator (\texttt{\#}) 
is encountered in a program expression.

\subsection{Installing Programs}

The predicate for downloading a page and installing its
corresponding LogicWeb program is given below. Its operation
parallels the stages shown in Figure 3.
\begin{alltts} 
download(Type, URL) :-
  created(Type, URL), !.                 % program already exists
download(Type, URL) :-                   % program does not exist
  retrieve(Type, URL, Contents),         % retrieve from the Web
  create_program(Type, URL, Contents),   % create the program
  assign_policyID(Type, URL, Contents).  % assign a policy program
\end{alltts}

If the program stored at \texttt{URL} is not already present
in the program store then its page is retrieved from
the Web, converted to a program, and assigned a policy.

The policy mechanism is located in \texttt{assign\_policyID/3},
which is defined as:
\begin{alltts}
assign_policyID(Type, URL, Contents) :-
  (pgpID_to_policyID(_, lw(Type, URL)) ->     % if policy program
    true         % lw(Type, URL) is not assigned a policy program
  ;
    determine_policyID(URL, Contents, PolicyID),
    record_policyID(Type, URL, PolicyID)
  ).
\end{alltts}

\begin{alltts}
determine_policyID(URL, Contents, PolicyID) :-
  (pgp_signed(URL) ->                    % if digitally signed
    authenticate(Contents, PGPID), % try authentication
    pgpID_to_policyID(PGPID, PolicyID)
  ;
    pgpID_to_policyID(unknown, PolicyID)
  ).

record_policyID(Type, URL, PolicyID) :-
  assert(policyID(lw(Type, URL), PolicyID)).
\end{alltts}

\begin{sloppypar}
\texttt{assign\_policyID/3} tries to obtain a security policy for
a downloaded (non-policy) LogicWeb program by calling 
\texttt{determine\_policyID/3}.
It bases its search on the result of calling \texttt{pgp\_signed/1} which
checks if the contents of the downloaded page are digitally signed. It
does this by inspecting the URL for the extension ``.lwpgp.html'',
which denotes a digitally signed LogicWeb program.
\end{sloppypar}

\begin{sloppypar}
If the program is signed then the signature is authenticated with
\texttt{authenticate/2}. It invokes PGP's authentication procedure
on \texttt{Contents}, which contains the HTML text and its digital
signature. This stage is depicted in Figure~3 in box B. PGP extracts
the digital signature from \texttt{Contents}, and attempts to decrypt
it to obtain an MD5 string by employing
its collection of public keys. These keys must have been previously
added to PGP by the LogicWeb administrator. On successful decryption,
PGP checks the HTML text from \texttt{Contents} against the MD5 string,
and returns the PGP ID labelling the key that decrypted the signature as
\texttt{PGPID}. On authentication failure (i.e., if the signature
is not decrypted or if the MD5 string does not match the HTML text),
\texttt{PGPID} is instantiated to \texttt{unknown}.
\end{sloppypar}

\begin{sloppypar}
The signature ID returned by \texttt{authenticate/2} is utilised
by \texttt{pgpID\_to\_policyID/2} to lookup a policy ID.
A typical \texttt{pgpID\_to\_policyID/2} fact:
\begin{alltts}
pgpID_to_policyID('Seng W. Loke <swloke@cs.mu.oz.au>',
           lw(get, "http://www.cs.mu.oz.au/\~{}swloke/my_policy.html").
\end{alltts}
The first argument is Seng Wai Loke's PGP ID,
and the second a policy program identifier.
\end{sloppypar}

\begin{sloppypar}
If the downloaded LogicWeb program is not signed then
\texttt{determine\_policyID/3} makes use of the default policy
ID associated with the `\texttt{unknown}' signature ID:
\begin{alltts}
pgpID_to_policyID(unknown,
      lw(get, "http://www.cs.mu.oz.au/\~{}swloke/default_policy.html").
\end{alltts}
\end{sloppypar}

\begin{sloppypar}
Back in \texttt{assign\_policyID/3}, the policy ID returned by
\texttt{determine\_policyID/3} is passed to \texttt{record\_policyID/3}.
It asserts a \texttt{policyID/2} fact which has as its first argument the
LogicWeb program ID and its second is the matching policy ID. This fact
can be used subsequently to map from a program ID to its policy. The
fact is not asserted into the LogicWeb program itself since it must be
impossible for a rogue program to redefine \texttt{policyID/2}.
\end{sloppypar}

\section{Control of Resource Usage}

This section considers denial-of-service attacks by showing how it is
possible to control an operation or resource with
policy programs and meta-interpreters.

\subsection{Resource Control Using Policy Programs}

\begin{sloppypar}
Resource usage can be monitored by keeping contextual or state information 
within a policy program, allowing decisions about the utilisation of
resources to be based on the execution history. For example, 
a limit can be imposed on the frequency of system calls. 

The following definition of \texttt{call\_system/1} permits
up to ten files to be opened at a time, thereby
limiting the number of allocated file descriptors.

\pagebreak
\begin{alltts}
call_system(P) :-                   % for non- open/3, close/1 calls
  P \(\backslash\)= open(_, _, _),
  P \(\backslash\)= close(_),
  built_ins:call_builtin(P).
call_system(open(F, R, S)) :-       % open a file
  open_count(N), N < 10,               
  increment_open_count,             % increment count of open files
  built_ins:call_builtin(open(F, R, S)). 
call_system(close(S)) :-            % close a file
  open_count(N), N > 0,               
  decrement_open_count,             % decrement count of open files
  built_ins:call_builtin(close(S)). 
\end{alltts}
\texttt{increment\_open\_count/0} and \texttt{decrement\_open\_count/0} 
update the value stored in \texttt{open\_count/1}.
\end{sloppypar}

\subsection{Resource Control Using Meta-interpreters}

\begin{sloppypar}
As pointed out in \cite{olw97},
denial-of-service attacks are not as severe as other kinds of
attack since the user can always hit the ``kill key'' and exit the
system. However, the graceful termination of goal evaluation is 
preferable so that the system and application can at least continue 
running. Execution control can be readily incorporated into the
system by using meta-interpreters~\cite{sb89,ss94}. Their use
for loop checking and detecting resource limits is described below.
\end{sloppypar}

\subsubsection{Loop Checking}

Our approach to loop checking uses two meta-interpreters.
\texttt{solve\_ad/2} evaluates a goal and also stores information 
about ancestor goals and the recursion depth.
Program~2 shows a simplified version of \texttt{solve\_ad/2} for 
pure Prolog.

\begin{program}
\begin{center}
\begin{alltts} 
\def\baselinestretch{1.0} 
{\footnotesize
solve_ad(_Depth_Ancs, true).         % evaluate true
solve_ad(Depth-Ancs, (G1, G2)) :-    % evaluate conjunction
  solve_ad(Depth-Ancs, G1),
  solve_ad(Depth-Ancs, G2).
solve_ad(Depth-Ancs, Goal) :-        % evaluate goal
  copy_term(Goal, CopyGoal),         % make a copy of the goal
  clause(Goal, Body),                % select a matching clause
  spy_point(Depth-Ancs, Goal),       % insert a spy_point
  Depth1 is Depth + 1,               % increment recursion depth
  solve_ad(Depth1-[CopyGoal|Ancs], Body).
}
\end{alltts}
\caption{A meta-interpreter for pure Prolog with an argument carrying the 
recursion depth and a list of ancestor goals.}
\end{center}
\end{program}
The purpose of the \texttt{spy\_point/2} call will be explained
shortly.

\begin{sloppypar}
The \texttt{solve\_t/2} meta-interpreter terminates goal
evaluation whenever some predefined termination condition becomes
true. Such conditions are specified as \texttt{terminate/1}
clauses. \texttt{solve\_t/2} also sets a termination flag to
`\texttt{terminated}' when a termination condition becomes true. Program~3
shows a simplified version of \texttt{solve\_t/2} for pure Prolog.
\end{sloppypar}

\begin{program}
\begin{center}
\begin{alltts} 
\def\baselinestretch{1.0} 
{\footnotesize
solve_t(true, _).         % evaluate true
solve_t(Goal, T) :- 
  terminate(Goal), !,     % check termination conditions
  T = terminated.      
solve_t((G1, G2), T) :-   % evaluate conjunction
  solve_t(G1, T), 
  (T == terminated -> 
     true
  ; 
     solve_t(G2, T)
  ).
solve_t(Goal, T) :-       % evaluate goal
  clause(Goal, Body), 
  solve_t(Body, T), 
  (T == terminated, ! ; true).
}
\end{alltts}
\caption{A meta-interpreter for pure Prolog which checks for
termination conditions.}
\end{center}
\end{program}

The next step is to combine these meta-interpreters with the LogicWeb
\texttt{demo/3} interpreter described in the previous section.
Instead of invoking \texttt{demo/3} directly, as in:
\begin{alltts}
  ?- demo([], empty, \(LogicWebGoal\)).
\end{alltts}
it will be called nested inside the two meta-interpreters:
\begin{alltts}
  ?- solve_t( solve_ad(0-[], demo([], empty, \(LogicWebGoal\))), _T ).
\end{alltts}

\texttt{solve\_ad/2} will record ancestor goals and recursion depth 
information about \texttt{demo/3}. \texttt{solve\_t/2} can check
for termination conditions by looking for suitable patterns in
the \texttt{spy\_point/2} calls in \texttt{solve\_ad/2}. For
example:
\begin{alltts}
terminate(spy_point(Depth-Ancs, demo(_, E, G))) :-
  (
    member(demo(_, E1, G1), Ancs),
    E = E1,               % the goals are in the same context  
    variant(G, G1),       % and they are variants of each other
    write('loop found')
  ;
    Depth > 40,           % recursion depth too large
    write('maximum recursion depth exceeded')
  ).
\end{alltts}
\texttt{demo(\_, E, G)} is the current LogicWeb interpreter goal,
\texttt{Depth} is the current recursion depth, and \texttt{Ancs} is the
list of ancestor goals. Note that the condition \texttt{E~=~E1} will not
detect loops where the context grows indefinitely. To detect such loops,
the size of the context can be compared against a preset limit.

The essential utility of this approach is the relative ease of combining
distinct meta-interpreters which monitor/control different aspects of the
computation. Other loop checking techniques are studied in \cite{bol91}.

\subsubsection{Resource Limits}

\begin{sloppypar}
The techniques from the last section can be easily
utilised to monitor resource limits. We will measure two 
resources: the number of LogicWeb programs downloaded, and
the number of clause applications. The first count will persist 
across different \texttt{solve\_t/2} invocations,
the second will be reset at each invocation of \texttt{solve\_t/2}.
\end{sloppypar}

The necessary \texttt{terminate/1} clauses for these measures are:
\begin{alltts}
terminate(invoked_builtin(download(_,_))) :-
  program_count(N),
  N > 100,           % permit up to 100 program downloads
  write('maximum LogicWeb program count exceeded').  
terminate(invoked_builtin(clause(_,_))) :- 
  retract(clause_count(N)),  
  N1 is N + 1,
  assert(clause_count(N1)),
  (N > 500 ->        % permit up to 500 clause applications
    writef('maximum clause count exceeded')
  ;
    !,
    fail
  ).
\end{alltts}
This code assumes that \texttt{solve\_ad/2} can deal with built-in
predicates, which would require a minor extension to Program~2.

The first \texttt{terminate/1} clause monitors program downloads by
looking for calls to the \texttt{download/2} built-in. When there is 
such a call, \texttt{program\_count/1} is checked to see if
the number of programs downloaded exceeds 100. If it does then execution
is terminated. The use of \texttt{program\_count/1} would require a
change to \texttt{create\_program/3} (called by \texttt{download/2})
which translates  a page into a program; it would also have to increment
\texttt{program\_count/1} each time a new program was created.

The second \texttt{terminate/1} clause restricts the number of
clause applications by monitoring calls to the \texttt{clause/2}
built-in. The \texttt{terminate/1} clause increments the number
stored in \texttt{clause\_count/1} until it exceeds 500.

\medskip

An oft-stated problem with nested meta-interpreters is the penalty
incurred upon execution speed. In fact, the termination checks described
here could be implemented more efficiently by directly modifying
\texttt{demo/3}. However, multiple meta-interpreters are less complex to
understand than a single, monolithic piece of code. Also, they permit a
compositional approach to implementing execution control: functionality
is separately implemented and introduced. We also believe that the
speed costs are relatively unimportant if the application carries out
frequent Web requests, and hence spends most of its wall-clock execution
time interacting with the Web. Moreover, partial reduction techniques
for translating meta-interpreters into specialised forms can be explored
to remove the levels of interpretation~\cite{ss94}.

\section{Comparisons with Security Models in Other Mobile Code Systems}

We review security models which share certain common features with
LogicWeb. First, we consider security approaches in some interpreted
languages, and then examine systems which utilise policy modules
and authentication.

\subsection{Security Models in Two Interpreted Languages}

\subsubsection{Safe-Tcl}
Tcl is an interpreted imperative language, and access to system
resources is via permitted commands of the interpreter. In the
Safe-Tcl security model~\cite{olw97}, security is enforced by
making dangerous commands unavailable to scripts running in the {\em safe
interpreter}. Potentially dangerous operations, such as opening sockets,
can be carried out via wrappers or aliases. The wrappers ensure that the
commands are used in a controlled manner (e.g., only socket connections
to some hosts are permitted).

\begin{sloppypar}
The security model in LogicWeb is partly motivated by the Safe-Tcl approach:
LogicWeb's valid system calls correspond to Safe-Tcl commands, and
\texttt{call\_system/1} corresponds to wrappers for those commands. A safe
interpreter in Safe-Tcl corresponds to the LogicWeb program interpreter
with appropriate policy programs. However, the Safe-Tcl security model
has no notion of authentication. 
\end{sloppypar}

Safe-Tcl allows programs running in different interpreters (each with
its own security policy) to communicate. Such communication effectively
composes the security policies of the programs, i.e. a program can use
the privileges of other programs. The dangers of this are
highlighted in~\cite{olw97}, but not solved in
a structured way. In LogicWeb, the composition of policies do not
violate safety criteria.

\subsubsection{Java Applets}
A Java applet is transmitted in bytecode format, and then
executed by an interpreter on the local host. This contrasts with
Safe-Tcl and LogicWeb where source code is transferred.

The LogicWeb system does not support concurrency or sophisticated
GUI programming, and so its security model does not need to deal
with multi-threading or screen resources (e.g., windows).  The Java
language enforces type safety by having the compiler ensure that class
methods do not access memory in ways that are inappropriate for their
type~\cite{secjava97}. LogicWeb programs are not explicitly typ\-ed
and there is no class abstraction. However, type conversions (e.g.,
strings to atoms) are done via system calls, and so the dangers of type
conversion are avoided at call time. There are two trust levels for
classes in the current Java security model: trusted classes are local
and part of the Java system, while untrusted classes are ones that are
downloaded. Multiple levels of trust are possible with signed
LogicWeb programs.

\begin{sloppypar}
Security for Java applets, as implemented in Microsoft's Internet
Explorer 3.0 and Netscape Navigator 3.0, disallows applets from
reading from or writing to local files, or establishing network connections
except to the originating host. However, there are ways around
these problems, including the use of server-side databases and
proxy servers. The digital signing of JAR (Java ARchive) files, as
supported in JDK 1.1.5\footnote{JDK 1.1.5 (final version) is available
at \texttt{http://java.sun.com/products/jdk/1.1/}.}, bundles Java code
and related files together into one file~\cite{secjava97}. In a similar
way, a collection of programs making up a LogicWeb application can be
bundled, signed, and transported as one archive file. 
\end{sloppypar}

\begin{sloppypar}
Microsoft Internet
Explorer 4.0 now allows finer grained control over the capabilities
granted to Java code, such as access rights to local files and network
connections~\cite{ie4security,ie4security2}. This is done by including a list of
capabilities requested by the applet in the applet's signature. Approval
for the capabilities are either pre-set or given via user dialog.
The implementation prevents transfer of privileges between classes
in a similar way to our model: if a class has not been directly granted a
capability, it cannot obtain that capability regardless of its caller's
capabilities. A class without a capability cannot gain that capability
by invoking a more privileged class. 
\end{sloppypar}

\subsection{Security Policy Modules in Two Mobile Code Systems}

\subsubsection{SERC's Safer Erlang (SSErl)}

The security mechanism in LogicWeb was motivated by the policy modules used
in SSErl, a functional declarative language for programming concurrent and
distributed systems~\cite{brown97a,brown97b}. In the SSErl framework, a
policy module is not associated with a particular program as in LogicWeb,
but with a SSErl node, a platform where multiple processes may run
concurrently. System operations are performed via built-in functions. A
policy module specifies the allowable built-in functions for all the
processes running at a node.

\begin{sloppypar}
In contrast to SSErl, execution of LogicWeb programs is single-threaded,
although the thread of execution can proceed from one program to another.
At any one time, at most one LogicWeb application is running. The overall
policy for the running application is represented by the current set of
policy programs, and is determined dynamically since it is generally
not possible to determine {\em a priori} which programs an application
will use.  
\end{sloppypar}

\subsubsection{Java Aglets}

\begin{sloppypar}
The Aglet workbench~\cite{klo97} allows Java programs (called {\em
aglets}) to be transferred between hosts as mobile agents.
A specialised language for writing aglet
security policy modules has been proposed which makes use of specific
roles such as the aglet's manufacturer, owner, execution platform (e.g.,
the URL of the host), and domain. In the LogicWeb implementation, the
assignment of policy programs is currently based solely on a program's
signature (as implemented by \texttt{determine\_policyID/3} in \S8.2). 
\end{sloppypar}

\subsection{Authentication in Two Mobile Code Systems}

\subsubsection{Agent Tcl}

Programs are authenticated with PGP in Agent
Tcl~\cite{gray96}. PGP is invoked externally in the same
way as in LogicWeb, thereby keeping the Agent Tcl system simpler and more
flexible (the encryption mechanism can be easily replaced). Depending
on the identity of the agent's owner, an Agent Tcl program is assigned
resources such as CPU time, windows, the file system, external programs,
and network connections.

\subsubsection{ActiveX}

{\em ActiveX controls}~\cite{ernst96} are programs (in native x86 code)
that can be executed on the client-side in Web pages. A digital
signature is attached to each ActiveX control which identifies the
control's author. The advantage of a signature is that since
the author is known, an ActiveX control is permitted more freedom to
access resources. 
 
Unlike a LogicWeb program, which is sandboxed by policy programs and
meta-in\-ter\-pre\-ters, a trusted ActiveX control is not restricted
in any way and has access to all operating system services. Since an
ActiveX control is native code, a suitable sandbox mechanism may be more
difficult to implement~\cite{omg}.

\section{Summary and Future Work}

Security is a major concern for any system that executes downloaded code.
This paper has described a security model for LogicWeb which protects
the client host from integrity, privacy, and denial-of-service attacks.
The security model is specified using proof rules in the operational
semantics of the language. This specification permits a straightforward
proof of the soundness of the model with respect to illegal system
operations. Resource control of predicates invoked at the
application level is carried out using policy programs. Resource control
of programs using resource counts and
loop checking is performed by meta-interpreters. Authentication 
provides varying trust levels, and different policy programs define
resource access for the different trust levels.

The security model dynamically extends the LogicWeb
system with selected policy programs. Since policy programs are separate
from the system (and integrated only on-demand) they are a convenient
means for imposing consistent security restrictions on multiple hosts.
For example, similar Intranet-wide security restrictions can be imposed
on a group of hosts by having them download the same policy programs
(which, if needed, can be specialised by being composed with host-specific
policies).

As noted in \cite{olw97}, it may be difficult to distinguish
between legitimate behavior and denial-of-service attacks. For instance,
it is easy to introduce resource bounds but difficult to state appropriate
values for those bounds. If a value is too low, a useful task may be
terminated prematurely; if too high, resource wastage occurs.  A possible
solution to this dilemma is to involve the user's judgment by providing
query-the-user facilities via a meta-interpreter. The user could also
utilise trace information supplied by the meta-interpreter to detect
unwanted behavior, such as the movement of data between sites using the
host's resources.

Some Prolog systems, such as SICStus Prolog, have a time-out 
facility for goal evaluation -- the goal is terminated if its
evaluation exceeds a given duration \cite{sicstus}. 
This can be incorporated as a hard limit after other checks in
our meta-interpreters:
\begin{alltts}
?- time_out(solve_t(solve_ad(0-[], demo([], empty, \(LogicWebGoal\))), _T), 
         Time, Result).
\end{alltts}
\texttt{time\_out/3} terminates the execution
of \texttt{solve\_t/2} after \texttt{Time} milliseconds, with 
\texttt{Result} being set to \texttt{timeout} or \texttt{success}.

Policy programs could be allocated based on the more general idea of
credentials instead of digital signatures alone. For instance, a program
might be allowed certain privileges not only because of its signature,
but because it had sufficient advocates. Recent work by Seamons {\em et
al}~\shortcite{sww97} on using Prolog to encode digital credential acceptance
policies and their verification logic could be exploited.

Policy programs simplify the administration of privileges to applications.
Our policy assignment depends solely on the digital signature
of the LogicWeb program. A more flexible method would involve 
multiple roles such as the program's manufacturer, owner, 
execution platform, or domain.
For instance, suppose we trust particular domains (or URLs) and 
assume that data transfer is not intercepted and modified.
We can assign policies based on the domains or URLs 
of the LogicWeb programs, regardless of whether they are signed. 
We modify \texttt{determine\_policyID/3} of \S8.2 as follows:

\begin{alltts}
safe("http://www.cs.mu.oz.au/~swl").            % trusted URLs
safe("http://www.cs.ait.ac.th/~ad").

determine_policyID(URL, _Contents, PolicyID) :- % safe URLs get a standard policy
  safe(URL), !,       
  PolicyID = lw(get, "http://www.policyforthetrusted.com").
determine_policyID(URL, Contents, PolicyID) :- % check signature for everything else
  (pgp_signed(URL) ->                    
    authenticate(Contents, PGPID), 
    pgpID_to_policyID(PGPID, PolicyID)
  ;
    pgpID_to_policyID(unknown, PolicyID)
  ).
\end{alltts}
Domain or zone-based policy assignment can be implemented this way.

\begin{sloppypar}
LW-com\-po\-si\-tion operators could be used to create
policies from dynamically composed programs.  
For instance, assume that the program at
\texttt{http://www.usedownloadedpgms.com} stores the clause:
\begin{alltts}
valid\_program(X) :- program_exists(X).
\end{alltts}
\texttt{determine\_policyID/3} can be modified to always
combine this policy with the one being downloaded:
\begin{alltts}
determine_policyID(URL, Contents, Policy) :- 
  (pgp_signed(URL) ->                    
    authenticate(Contents, PGPID), 
    pgpID_to_policyID(PGPID, PolicyID)
  ;
    pgpID_to_policyID(unknown, PolicyID)
  ),
  Policy = lw(get,"http://www.usedownloadedpgms.com")+PolicyID. % combine policies 
\end{alltts}
\end{sloppypar}

We have only considered policy programs created by the user of the
LogicWeb system. If policies could be created by other parties,
authentication would then become necessary for policy programs in
addition to ordinary LogicWeb programs.

A limitation of our model is that the system must contain all the
public keys required for the authentication of incoming programs.
An automatic distribution mechanism for PGP public keys is needed.
For example, the LogicWeb system could extract required public
keys from homepages given their URLs, or could query Internet PGP
public key servers\footnote{A PGP public key server is located at\\
\texttt{http://www-swiss.ai.mit.edu/\~{}bal/keyserver.html}.}~\cite{garfinkel95}.
But this requires that the information be trusted, has not been tampered
with, and is reliable. The use of credentials, i.e.\  certification by
other authorities, would be required.

The LogicWeb security model is conservative in that a trusted program
is not allowed to transfer its privileges to an untrusted program it
is using. Besides resource usage privileges, another kind of privilege
could be introduced: the right to transfer privileges. This
would be useful when the system did not know the signatories of all the
programs that a trusted program utilised. The policy program would have to
specify which privileges were transferable, and whether the right to
transfer privileges was itself transferable.

\bibliographystyle{acmtrans}

\end{document}